\documentclass[fleqn,usenatbib]{mnras}

\usepackage[T1]{fontenc}

\DeclareRobustCommand{\VAN}[3]{#2}
\let\VANthebibliography\thebibliography
\def\thebibliography{\DeclareRobustCommand{\VAN}[3]{##3}\VANthebibliography}

\usepackage{graphicx}
\usepackage{subcaption}
\usepackage{amsmath}
\usepackage{amssymb}
\usepackage{url}
\usepackage{enumitem} 

\usepackage{subfiles}
\graphicspath{{Images/}{../Images/}} 

\newcommand{\subbib}{\bibliography{PhD, Additional_bibliography}} 

\usepackage[normalem]{ulem} 
\usepackage{wasysym}

\newcommand{\ie}[0]{$\textnormal{i.e. }$}
\newcommand{\tn}[1]{\textnormal{#1}}
\newcommand{\sub}[1]{_{\textnormal{#1}}}
\newcommand{\Msun}[0]{\,\textnormal{M}_{\textnormal{\astrosun}}}

\newcommand{\logMm}[0]{\textnormal{log}_{10}(M_{*}/\Msun)}

\newcommand{\oii}{[\ion{O}{ii}]}
\newcommand{\oiii}{[\ion{O}{iii}]}
\newcommand{\oiiiauroral}{[\ion{O}{iii}]$\lambda$4363\ }
\newcommand{\siii}{[\ion{S}{iii}]}
\newcommand{\nii}{[\ion{N}{ii}]}
\newcommand{\feii}{[\ion{Fe}{ii}]$\lambda$}

\usepackage{newtxtext,newtxmath}

\title[Investigating Central Dips in Metallicity]{Investigating the Origin of Observed Central Dips in Radial Metallicity Profiles}

\author[B. Easeman et al.]{
Bethan Easeman,$^{1}$\thanks{E-mail: be329@bath.ac.uk}
Patricia Schady,$^{1}$
Stijn Wuyts$^{1}$
and Robert M. Yates$^{2}$
\\
$^{1}$Department of Physics, University of Bath, Bath, BA2 7AY, United Kingdom\\
$^{2}$Department of Physics, University of Surrey, Surrey, GU2 7XH, United Kingdom\\
}

\date{Accepted XXX. Received YYY; in original form ZZZ}

\pubyear{2021}

\begin{document}
\label{firstpage}
\pagerange{\pageref{firstpage}--\pageref{lastpage}}
\maketitle

\renewcommand{\subbib}{}

\begin{abstract}

Radial metallicity trends provide a key indicator of physical processes such as star formation and radial gas migration within a galaxy. Large IFU surveys allow for detailed studies of these radial variations, with recent observations detecting central dips in the metallicity, which may trace the impact of various evolutionary processes. However, the origin of these dips has not been conclusively determined, with suggestions that they may be diagnostic dependent. In this paper, we use the SDSS-IV MaNGA survey to investigate whether the observed dips represent genuine decreases in the central metallicity, or if they could be an artefact of the diagnostic used. Using a sub-sample of 758 local star-forming galaxies at low inclinations, we investigate in detail the impact of using different strong line diagnostics on the shapes of the returned profiles, and the prevalence of dips. We find no clear evidence of the dips being caused by changing values of the ionisation parameter within galaxies. To investigate physical causes, we explore both global and spatially-resolved parameters, finding that galaxies exhibiting central dips in the O3N2 metallicity profile have on average lower H$\alpha$EW values out to $R/R\sub{e} \sim 1.5$, and higher values of D$_N$(4000) in the central regions. We additionally find a higher prevalence of dips in galaxies with high stellar mass, and lower values of global specific star formation rate, suggesting a possible link to central quenching. Nevertheless, these results are dependent on the diagnostic used, suggesting caution should be taken when interpreting observed features in galaxy metallicity gradients.

\end{abstract}



\color{black}

\section{Introduction}
\label{sect:Intro}

Metals are formed, and later dispersed into the surrounding interstellar medium (ISM), during the life cycles of stars. The gas-phase metallicity of a galaxy therefore provides a key tracer for the history of star formation within the galaxy, as well as providing evidence for evolutionary processes such as inflows or outflows of gas. This allows measurements of the metallicity, especially from spatially-resolved observations, to test, inform, and provide important constraints on models of galactic formation and evolution.

The metallicity gradient, \ie{}the slope of the radial metallicity profile, provides information on the overall radial distribution of the gas-phase metallicity, allowing for study of radial variations in physical processes occurring within the galaxy. The inside-out model of galaxy growth, presented by \cite{Matteucci1989}, suggests that gas should reach sufficient density for star formation first in the central regions of galaxies, leading to higher levels of chemical enrichment in these regions, built up by successive generations of stars. Star formation is then expected to progress to the outer regions over time. This suggests that smooth negative metallicity gradients should therefore be observed, with metallicity decreasing towards the outer regions. Metallicity gradients have therefore been studied to test these predictions, with extensive studies of the metallicity gradient within our own Galaxy \citep[e.g.][]{Shaver1983, Pilyugin2003, Esteban2017} and using multi-slit observations of external galaxies to study the effects of evolutionary processes such as mergers \cite[e.g.][]{KennicuttJr.2003, Kewley2010, Kudritzki2015, Bresolin2019, Esteban2020}. 

Modern integral field unit (IFU) spectroscopy has revolutionised these studies, producing 2-dimensional maps of galaxies and allowing the variation of metallicity across the entire galaxy to be studied out to typically $\sim$1-2.5 effective radii ($R\sub{e}$). With such spatially-resolved spectroscopic data it has been possible to observe additional features beyond the approximation of a single linear metallicity gradient, showing that a number of galaxies exhibit notable deviations such as central dips, or a flattening in the outer regions \citep[e.g.][]{Sanchez2014, Sanchez-Menguiano2018}, building upon the first suggestions of central dips presented in \cite{Belley1992}. Due to the complexity of obtaining these measurements, and the complicated nature of determining the impact of various evolutionary processes on the metallicity content of the ISM, no definitive conclusion has been reached on the processes causing these features. Links between the presence of these features and global properties of the galaxies have been explored, with multiple works finding central dips becoming increasingly apparent with increasing global stellar mass \citep{Belfiore2017, Sanchez-Menguiano2018, Schaefer2019, Mingozzi2020, Yates2021}. The effect of bars within spiral galaxies causing increased radial mixing and consequently flattened inner metallicity profiles has also been considered \citep{Zurita2021}, while other works have found no clear link between galaxy morphology and the presence of central metallicity dips \citep{Sanchez2014}. Inflowing or outflowing gas has also been explored as a potential mechanism, with radial motions of gas suggested as a possible cause of the observed metallicity dips \citep{Rupke2008, Kewley2010, Sanchez2014, Sanchez-Menguiano2016}. Inflowing pristine gas has been suggested as a mechanism to dilute the metallicity within the central regions \citep[e.g.][]{Kewley2010, Rupke2008, Mannucci2010}, and outflows are also known to drive metal-rich gas from the central regions \citep[e.g.][]{Tremonti2004}.

A link between the global stellar mass and the metallicity of a galaxy has long been established \citep[e.g.][]{Lequeux1979, Tremonti2004}, with more recent works using IFU data finding that this mass-metallicity relation (MZR) persists on spatially-resolved scales \citep[e.g.][]{Sanchez2013, Cano-Diaz2016, Ellison2018}. A number of works have therefore explored whether there is any link between the metallicity gradient within a galaxy, and its global stellar mass. However, contradictory results have been found, with some works finding a correlation (at least at lower masses) to exist \citep[e.g.][]{Belfiore2017,Poetrodjojo2018,Franchetto2021}, others an anti-correlation \citep[e.g.][]{Kaplan2016,Erroz-Ferrer2019,Yates2021}, and some finding no correlation \citep[e.g.][]{Sanchez2014, Lian2018}. Some works have found that evidence for a correlation is dependent on whether the gradients are normalised by a characteristic radius, such as the galaxies' $R\sub{e}$ rather than measured in physical units of dex/kpc \citep[e.g.][]{Ho2015, Sanchez-Menguiano2016,Bresolin2019}.

Another complication arises from the diagnostic tools used to measure the metallicity. Methods relying on measuring the electron temperature ($T\sub{e}$) of the gas, and determining the metallicity using the known anti-correlation between the two parameters, are widely considered to provide the most accurate and reliable results. However, these methods require the detection of auroral lines, commonly the \oiiiauroral line, which are very faint within optical spectra - the flux of the \oiiiauroral line is typically $\sim$100 times fainter than that of the \oiii $\lambda$5007 line \citep{Schaefer2019}. This effect is worsened at higher metallicities due to the increased cooling of gas via the metal lines in these metal-rich regions \citep{Hoyos2006}. The $T\sub{e}$ method may also intrinsically under-estimate the true metallicity in metal-rich and/or low-ionisation environments \citep[e.g.][]{Kobulnicky1999,Stasinska2005,Kewley2008,Yates2020}.

Diagnostics which calibrate ratios of stronger emission lines to $T_{\rm e}$-based metallicities or photoionisation models have therefore been developed for use when the auroral lines cannot be reliably detected. These strong line diagnostics are not without their complications, with certain diagnostics having strong dependencies on the ionisation parameter \citep{Kewley2002}, which can cause the diagnostics to break down on sub-HII region scales \citep[e.g.][]{Kruhler2017,Mao2018}, for example. Systematic offsets between strong line diagnostics are also known to exist. For example, diagnostics calibrated using photoionisation models typically return higher metallicity values than empirically derived methods by $\sim$0.4-0.6 dex \citep{Kewley2008, Teimoorinia2021}. It is also important for the sample of study to have similar physical characteristics to the original calibration sample when using strong-line diagnostics \citep[e.g.][]{Stasinska2010}. In addition, strong optical emission lines saturate at low electron temperatures, when free electrons no longer have enough energy to easily excite the ions into producing these lines. This leads to the double-valued nature of some diagnostics within the range of metallicities typically seen in H\textsc{ii} regions \citep[e.g.][]{Kewley2002}. Finally, the measured gradients have also been found to vary when different strong line diagnostics are used \citep{Erroz-Ferrer2019, Sanchez-Menguiano2018, Cameron2020, Poetrodjojo2021,Yates2021}. 

Given the dependence of measured galaxy metallicity gradients on the strong line diagnostic used, and the important implications that features such as central dips have on galaxy evolution models, the aim of our analysis is to investigate whether the observed central metallicity dips represent genuine decreases in the metallicity in the central regions, or if they can be explained as an artefact of the diagnostic used. To investigate potential physical causes of the dips, we explore the link between galaxies exhibiting central dips, and both global and spatially-resolved physical parameters. 

In studies of gas-phase metallicity, the relative abundance of oxygen is commonly used as a proxy, as oxygen is the dominant metal by mass, and has bright, easily observed optical emission lines \citep{Kewley2008}. We follow this convention, and throughout this paper, the term metallicity will refer to the gas-phase oxygen abundance, expressed as 12+log(O/H).  

In Section \ref{sect:ObsData} we describe the selection of our galaxy sample from the MaNGA DR15 data. In Section \ref{sect:Zmaps} we discuss the process of obtaining or deriving flux maps of the emission lines required for our analysis, as well as implementing the various diagnostics to produce maps of the metallicity and ionisation parameter. Our method of fitting and categorising the radial metallicity profiles is also covered here, and our results are presented in Section \ref{sect:Results}. In Section \ref{sect:logU} we explore whether the dips could be explained as an artefact resulting from a changing relationship between ionisation parameter and metallicity in the central regions of galaxies. We then investigate whether the galaxies exhibiting central dips show differences from the rest of the sample in any global parameter space, or in spatially-resolved average profiles of various parameters in Section \ref{sect:GalaxyProperties}. Our discussion of the results is in Section \ref{sect:Discussion}, and we present our conclusions in Section \ref{sect:Conclusions}.

\subbib

\section{Selection of Observational Data}
\label{sect:ObsData}

\subsection{MaNGA Survey}

We selected galaxies from Data Release 15 of the MaNGA survey \citep[Mapping Nearby Galaxies at Apache Point Observatory;][]{Bundy2015}, part of the SDSS-IV project dedicated to obtaining IFU spectroscopy for 10,000 galaxies within the local universe ($0.01 < z < 0.15$) \citep{Blanton2017}. The galaxies observed by MaNGA were selected without cuts on inclination, morphology, size, or environment, and the sample was designed such that there is an approximately flat global stellar mass distribution \citep{Wake2017}. Observations are made using the 2.5 m telescope at Apache Point Observatory \citep{Gunn2006}, and the BOSS spectrograph \citep{Smee2013, Drory2015}, with galaxies from the Primary and Secondary samples observed out to 1.5 $R\sub{e}$ and 2.5 $R\sub{e}$, respectively. MaNGA has a wide wavelength range of 3600-10300\AA{} at R$\sim$2000 \citep{Bundy2015}, covering most of the emission lines commonly used in gas-phase metallicity diagnostics.

\subsection{Sample Selection}\label{sec:sampleselection}

\begin{figure}
\centering
\begin{subfigure}{1\columnwidth}
\centering
\includegraphics[width=0.8\columnwidth]{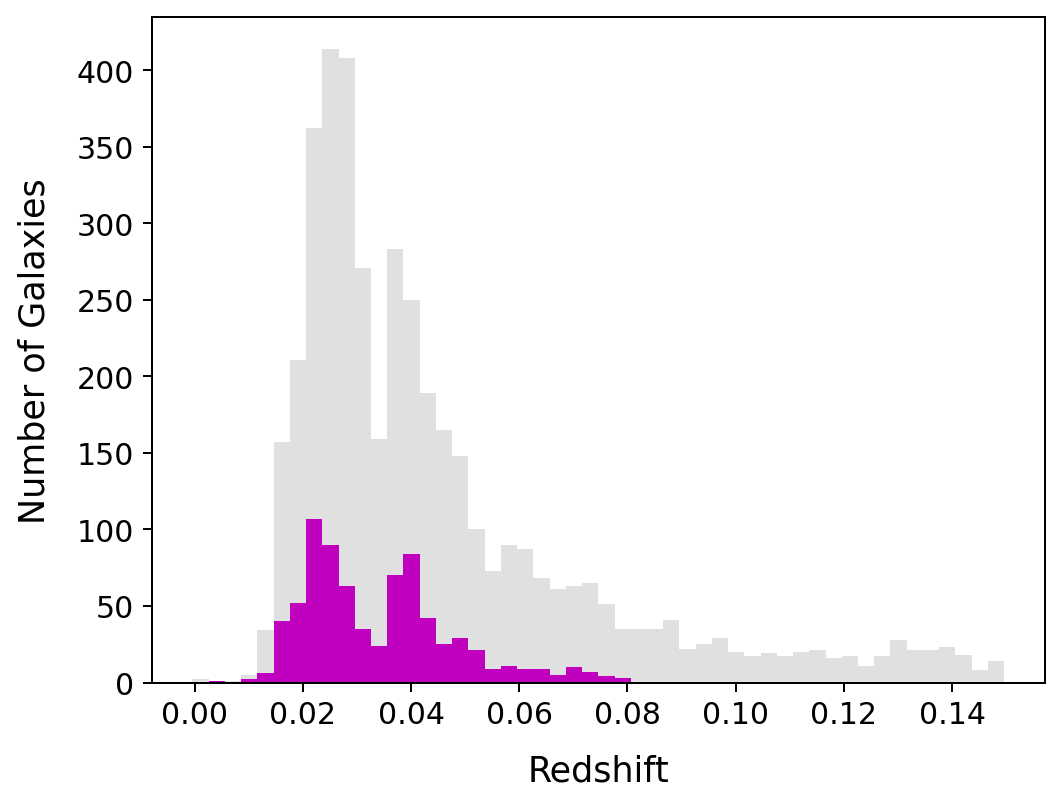}
\end{subfigure}  %
\begin{subfigure}{1\columnwidth}
\centering
\includegraphics[width=0.8\columnwidth]{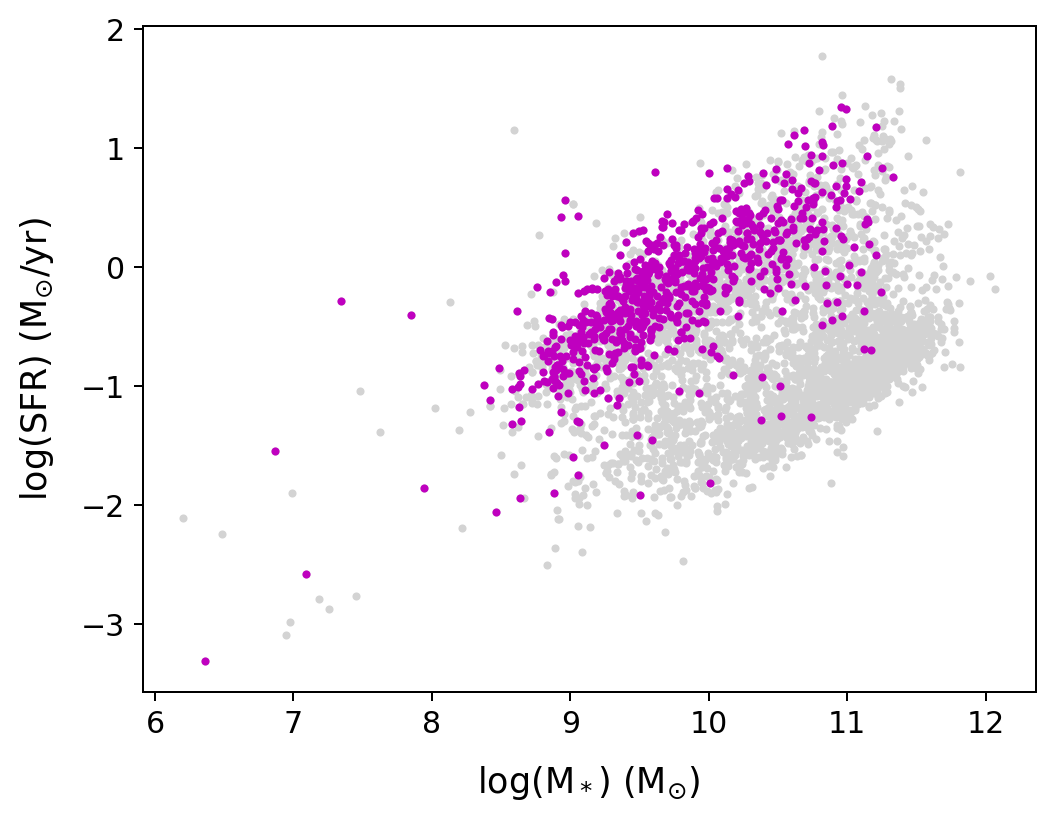}
\end{subfigure}
\caption{The parent sample of 4248 galaxies is shown in grey, and our selected sample of 758 galaxies is overlaid in magenta. \textit{Top}: The redshift distribution of the galaxies is seen to follow that of the parent sample, with a bimodal distribution owing to the Primary and Secondary MaNGA samples. \textit{Bottom}: As we required galaxies to be star-forming according to the BPT diagram, our sample can be seen to lie predominantly in the upper locus of galaxies in the main sequence parameter space, associated with star-forming galaxies. The lower population of passive galaxies can be clearly seen in the parent distribution.} 
\label{fig:SFR_Mstar_redshift}
\end{figure} 

The initial sample of galaxies were taken from DR15 of the MaNGA survey \citep{Bundy2015}, containing observations of 4688 galaxies, and matched to the MPA-JHU catalogue to provide the global stellar mass and star formation rate for each galaxy \citep{Kauffmann2003b, Brinchmann2004, Tremonti2004, MPAJHU}. We note that the global stellar mass values reported in the MPA-JHU catalogue may be underestimated for galaxies below $\logMm{}\sim{}10.0$, compared to more recent analysis of SDSS photometry for the NASA-Sloan Atlas (NSA) catalogue \citep{Blanton2011, Yates2020}, see further discussion in Section \ref{subsect:GlobalMassSFR}. The global properties of the galaxies will be used to explore whether there is any link between the fitted metallicity profiles and any physical properties of the galaxies. Out of these galaxies, 4248 could be matched to the MPA-JHU catalogue, so these were taken as the full parent sample. 

Using the MaNGA Data Analysis Pipeline (DAP) quality flags, we removed any galaxies with flags indicating issues other than foreground stars, as we found the foreground stars to be well-covered by the mask maps used throughout the rest of our analysis to remove any problematic individual spaxels. We then selected our sample of galaxies using a combination of cuts on global and spatially-resolved parameters. 

As the strong line metallicity diagnostics were calibrated against emission from gas ionised by hot young stars in HII regions, care must be taken to remove galaxies from our sample which have significant emission from other sources, such as an active galactic nucleus (AGN). To achieve this, we removed galaxies from the sample if the line ratios from the global spectrum (determined using the summed IFU flux maps) placed the galaxy outside of the star forming (SF) region of the Baldwin, Phillips \& Terlevich \citep[BPT; ][]{Baldwin1981} diagnostic diagram, as defined by the \cite{Kauffmann2003} demarcation line. To ensure that the metallicity gradient of the galaxy could be reliably determined, we additionally required the galaxy to be relatively face-on (projected minor-to-major axis ratios, $b/a$, $> 0.6$, corresponding to inclinations of less than 53 degrees when assuming no intrinsic ellipticity), and to have an effective radius, R$_{e}$, greater than 5". 

One of the physical properties that we later explore is the ionisation parameter, \textit{U}, and how it may affect the shape of the measured metallicity profile when using different diagnostics. The ionisation parameter describes the ability of a source to ionise the surrounding gas, and can be considered the speed at which a source can drive an ionisation front through neutral gas  \citep{Kewley2019}. To measure the ionisation parameter we chose to use a diagnostic based on the ratio of single-to-double ionised sulphur (rather than the more metallicity-dependent oxygen-based diagnostics),  which sets a redshift limit on our sample of $z<0.08$ in order to ensure coverage of the \siii$\lambda$9531 line. 

Finally, following the process used by \cite{Schaefer2019}, galaxies were removed from the sample if less than 60\% of the spaxels with $g$-band signal/noise (S/N) $>$ 2 were determined to give reliable metallicity measurements. Spaxels were deemed to give reliable measurements if they met the following criteria:
\begin{enumerate}[leftmargin=*]
\item The spaxels must fall within the SF region of the \nii\ BPT diagram (below the \cite{Kauffmann2003} line). 
\item They must have fluxes for the \oiii$\lambda$5007, \nii$\lambda$6585, H$\alpha$ and H$\beta$ lines,  commonly used in strong-line metallicity diagnostics, measured with  S/N $>$ 3.
\item We additionally required that the spaxels must have H$\alpha$ equivalent width (H$\alpha$EW) $>$ 6 \AA, to ensure the spaxels were not significantly contaminated by emission from Diffuse Ionised Gases (DIGs). 
\end{enumerate}

Emission from DIGs can cause issues when applying strong line metallicity diagnostics due to the differing physical conditions of DIGs compared to the gas in the HII regions used to calibrate the diagnostics. The ionising source of DIGs remains uncertain, but the physical conditions are known to vary widely from those within HII regions, with DIGs having lower densities and ionisation parameters \citep{Sanders2017, Zhang2017, Lacerda2018a}. Consequently, results from applying strong line diagnostics to spectra contaminated by emission from DIGs are considered unreliable \citep[e.g. ][]{Zhang2017}. The MaNGA observations have a $\sim$kiloparsec scale resolution \citep{Wake2017}, therefore it is not possible to entirely isolate HII regions, which are typically on the order of 100 pc in size, from regions of DIGs. Instead, we aim to remove spaxels which show significant contamination from DIGs from our analysis. 

A higher value of 14 \AA\ was suggested by \cite{Lacerda2018a} to separate out purely star-forming regions from any pure DIGs or composite regions, however, we found that when implementing such a conservative approach, we removed a large fraction of spaxels which lay clearly in the SF region of the BPT diagram. To minimise the fraction of spaxels removed from our galaxy data cube, while still limiting the contamination from DIGs emission, we therefore chose to use a less conservative cut on H$\alpha$EW of $>$ 6 \AA\, combined with removing any spaxels that lay in the composite or LIER region of the BPT diagram.

These requirements left us with a final sample of 758 galaxies, the redshift distribution and location of these galaxies on the star-forming main sequence, compared to that for the parent sample, can be seen in Fig.~\ref{fig:SFR_Mstar_redshift}. The two peaks seen in the redshift distribution correspond to the Primary (covered out to 1.5 R$_{e}$) and Secondary (covered out to 2.5 R$_{e}$) MaNGA samples \citep{Yan2016}, we verified that this bimodal distribution did not affect the results we obtain in our analysis.

Galaxy morphology has been suggested to impact the obtained radial metallicity profiles \citep{Yates2021}, with non-disc/spiral galaxies tending to have flatter gradients. We made no cut on morphology when selecting our sample as we wish to examine trends in metallicity profile for all galaxies, rather than just those with structured discs.

\subsection{Low-luminosity AGN}
When selecting the sample of galaxies, our method of selecting SF galaxies using the summed IFU flux maps and the requirement for 60\% of the spaxels to give reliable metallicity measurements may not remove any galaxies with centrally concentrated low-luminosity AGN. To check whether our sample had significant contamination from such galaxies, we used stacked spectra from the central region ($\leq0.25 R\sub{e}$) of galaxies, to place the galaxies on the BPT diagram \citep[see][for details]{Avery2021}. These spectra placed 674 (89\%) of our galaxies within the SF region denoted by the \cite{Kauffmann2003} line, and only 3 of our 758 galaxies indeed hosted central low-luminosity AGN. Most of the non-SF galaxies were categorised as composite galaxies using this method (63 galaxies), with 18 galaxies falling in the LIER region \cite{Kewley2006}. We verified that these non-SF galaxies are not concentrated within one particular metallicity profile category, as introduced in Section 3.2, suggesting that our results would not be significantly altered had we applied this more stringent condition in our sample selection.

\subbib

\section{Methodology}
\label{sect:Zmaps}

To study the shapes of the metallicity profiles of the galaxies within our sample, and explore any dependencies on the strong line diagnostic used, and on the galaxies' physical properties, we made use of the MAPS files provided by the MaNGA DAP \citep{Westfall2019}. We used the data derived using the hybrid binning scheme, as recommended by \cite{Belfiore2019} for emission-line science. The hybrid binning scheme also maximises our spatial resolution, which is preferable since lower spatial resolution can cause flattening of observed gradients \citep[e.g.][]{Acharyya2020, Poetrodjojo2019}. The MAPS files include 2D maps of the emission line fluxes, corrected for Galactic reddening \citep{Belfiore2019}, and maps of the radial distance of each spaxel from the centre of the galaxy.

Flux maps for all of the emission lines required in the strong line metallicity diagnostics we used (as detailed in Section \ref{subsect:Diagnostics}) are available within the MAPS files. However, maps for the \siii$\lambda\lambda$9070, 9531 lines were not included in the DAP, due to the adopted MILES stellar library used to separate the stellar and gas components in the spectra being limited to below $\sim$7400 \AA\ \citep{Belfiore2017}. We therefore produced these \siii\ maps ourselves following a similar process to that used within the DAP \citep{Belfiore2019}, as described in Section \ref{subsect:IonParamMaps}.

\subsection{Metallicity Diagnostics Considered}
\label{subsect:Diagnostics}

We used a variety of strong line diagnostics to gain a greater understanding of the effects of using different diagnostics on the shape of the derived metallicity profiles. The diagnostics we used are: 

\begin{enumerate}[leftmargin=*]
    \item the \cite{Curti2017} N2 and O3N2 diagnostics, chosen as they are commonly used, and known to be dependent on the ionisation parameter \citep[e.g.][]{Kewley2002, Pettini2004}. We chose to use the \cite{Curti2017} re-calibration of these diagnostics, as it is a recent calibration, using updated atomic data as well as a large galaxy sample covering a wide range of metallicity values. 
    \item the \cite{Kobulnicky2004} R23 diagnostic, as it explicitly accounts for the dependence of 12+log(O/H) on $\log(U)$. We chose to use the \cite{Kobulnicky2004} parameterisation of this diagnostic, as it is widely used, and the \cite{Curti2017} re-calibration was not calibrated to lower metallicities due to the complication of the double-branched nature. Whether log([\ion{N}{ii}]/\oii) returned a value above or below 1.2 was used to determine whether the upper or lower branch was to be used \citep{Kewley2008}.
    \item the \cite{Dopita2016} N2S2H$\alpha$ diagnostic (hereafter, the D16 diagnostic), as they argue that the dependence of this diagnostic on the ionising conditions within HII regions is negligible, so if a changing relationship between the metallicity and $\log(U)$ were to be the cause of dips, this diagnostic should provide a useful benchmark to test against. 
\end{enumerate}

Before applying the strong line diagnostics to produce 2D maps of the metallicity, the flux maps provided from the DAP needed to be corrected for attenuation within the host galaxy. To do this, we measured the Balmer decrement (H$\alpha$/H$\beta$), and then applied the \cite{Calzetti2000} dust attenuation law, as used by \cite{Belfiore2017}, to determine the extinction at the wavelength of each emission line. \cite{Belfiore2017} found that using the \cite{Cardelli1989} extinction law instead produced similar results. We assumed an intrinsic Balmer decrement of 2.87 \citep{Osterbrock2006}, suitable for star forming galaxies with electron density $\sim$100cm$^3$ and temperature $\sim$10,000K.  

With these dust-corrected flux maps, we then used the various diagnostics to produce 2D maps of the metallicity. In order to remove any spaxels which may return unreliable values for the metallicity, we first removed any spaxels which show significant contamination from DIGs, as the presence of DIGs has been shown to significantly bias both absolute metallicity values, as well as metallicity gradients \citep{Zhang2017}. We used the H$\alpha$EW, which has been shown to trace the presence of DIGs \citep{Lacerda2018a}, to determine which spaxels were significantly contaminated, removing any spaxels with H$\alpha$EW $<$ 6 \AA{} (see Section~\ref{sec:sampleselection} for further discussion on DIGs). 

Similar to our selection criteria for the galaxy sample described in Section~\ref{sec:sampleselection}, we combined this H$\alpha$EW cut with a requirement for the spaxels to fall within the SF region of the BPT diagram, below the \cite{Kauffmann2003} demarcation line. This was done to ensure any spaxels with contamination from alternative ionisation sources such as an AGN were removed. Additionally, contamination from DIGs has been shown to move emission from HII regions towards the LI(N)ER-like region of the BPT diagrams \citep{Zhang2017}, therefore this cut should act alongside the H$\alpha$EW cut to remove any remaining DIGs-contaminated spaxels. Finally, any spaxels with S/N $<3$ in the lines required for each diagnostic were removed, to ensure the metallicity could be reliably determined. 

We investigated whether it was possible to determine radial metallicity gradients using $T\sub{e}$-based metallicity diagnostics, identifying a sub-sample of 33 galaxies with S/N of the \oiiiauroral{} line > 3 in the global spectrum. We stacked spectra within radial annuli, in order to increase the S/N of the \oiiiauroral{} line. However it was not possible to detect the line with sufficient S/N within a sufficient number of radial annuli to constrain the shape of the metallicity gradients.

\subsection{Fitting and Classification of Metallicity Profiles}
\label{subsect:FittingProfiles}

\begin{figure*}
\centering
\includegraphics[width=1\linewidth]{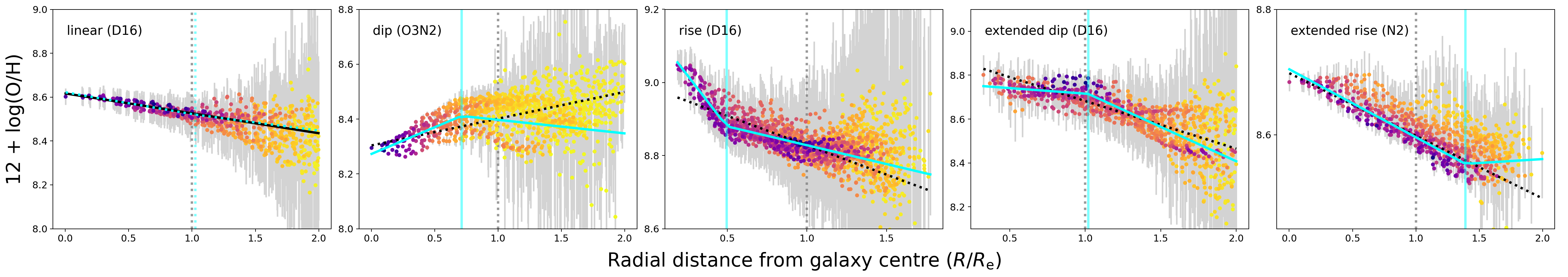}
\caption{Examples of metallicity profiles categorised by a range of metallicity diagnostics as linear, central dip, central rise, extended dip, and extended rise, respectively. The two fitted models can be seen as the black and cyan lines, with the broken power law plotted in cyan, and the single power law in black, with the solid line being the preferred model according to the AIC. The data points are coloured purple-to-yellow by 1/$\sigma$. The vertical cyan line marks $R\sub{br}$ for the broken power law, and the vertical grey line marks 1 $R\sub{e}$. Note that these profiles have been selected as giving clear examples of each category, Fig.~\ref{fig:avgMet} shows the wide variation in the profiles obtained across the sample of galaxies. The masses of the galaxies shown are 9.6 M$_\odot$, 9.0 M$_\odot$, 10.7 M$_\odot$, 10.3 M$_\odot$, and 9.7 M$_\odot$, respectively.}
\label{fig:MetProfiles}
\end{figure*}

After producing these 2D maps of metallicity, and selecting out only the spaxels which would provide us with reliable measurements, we combined these with maps of the radial distance between each spaxel and the centre of the galaxy, deﬁned along elliptical isophotes to account for projection effects, and normalized to the elliptical Petrosian $r$-band $R\sub{e}$ provided by the NSA catalogue \citep{Westfall2019}. This allowed us to produce radial profiles of the metallicity within the galaxy, and examples of these profiles can be seen in Fig.~\ref{fig:MetProfiles}. Uncertainties were taken from converting values from the inverse variance maps in the MAPS files to standard deviations ($\sigma$), and were propagated throughout the calculations using the \texttt{uncertainties} python package \citep{PythonUncertainties}. 

In order to quantify the shapes of these radial profiles, we fitted each profile with two different models. We fit the profiles with both a single linear power law model, and with a broken power law model, to explore whether the profiles showed evidence for radial features such as central dips. We used the \texttt{curve\_fit} function from the optimize module of SciPy to perform these fits \citep{SciPy}, taking into account our 1-sigma metallicity uncertainties to apply a least-squares minimisation. We removed any spaxels at distances $>$2 $R\sub{e}$ from the galaxy centre, as we found that including these spaxels did not help to further constrain the fits. 

To determine which of these two models provided the best description of the data, we used the Akaike Information Criterion \citep[AIC;][]{Burnham2002}, given by
\begin{equation}
AIC = 2k + n\tn{ln}(RSS)\;\;.
\label{eqn:AIC}
\end{equation} 

This criterion compares the goodness of fit, quantified by the residual sum of squares (\textit{RSS}), to the complexity of the model, quantified by the number of parameters (\textit{k}), to determine which of two models provides a statistically significantly better fit.

The model with the smallest AIC value ($AIC\sub{min}$) was taken as the preferred model if the probability ($P$) of the alternative model providing an equal or better description of the data is $<0.05$, 
\begin{equation}
P = \tn{exp}((AIC\sub{min} - AIC\sub{max})/2)\;\;,
\label{eqn:AICp}
\end{equation}  

where $AIC\sub{max}$ is the AIC value of the alternative model. In the case where $P>0.05$, there is no preferred model, and the best-fit model is marked as `inconclusive'. For galaxies where the cuts applied left no spaxels within 0.5 R$_{e}$, the fitted model was also marked as `inconclusive'; this only affected a small number of galaxies, typically around $\sim$1\% for most of the applied metallicity diagnostics, reaching a maximum of 5\%. 

We found that our sample of galaxies could be well-described by 5 categories based on the properties of the best-fit model. Aside from power law and broken power law profiles, we additionally considered the radial position of the break ($R\sub{br}$), as well as the relationship between the gradients at either side of the break (defined as $\alpha_{in}$ and $\alpha_{out}$). Galaxies best-fit by a single power law according to the AIC were categorised as having `linear' profiles, and in the case where a broken power law provided a better fit, the galaxies were categorised as defined below.  

\begin{itemize}[leftmargin=*]
\item (central) dip: $R\sub{br}\leq1 R\sub{e}$,\ \ $\alpha\sub{in}>\alpha\sub{out}$.
\item (central) rise: $R\sub{br}\leq1 R\sub{e}$,\ \ $\alpha\sub{in}<\alpha\sub{out}$.
\item extended dip: $R\sub{br}>1 R\sub{e}$,\ \ $\alpha\sub{in}>\alpha\sub{out}$.
\item extended rise: $R\sub{br}>1 R\sub{e}$,\ \ $\alpha\sub{in}<\alpha\sub{out}$.
\end{itemize}

Examples of metallicity profiles and the fitted models for each of these categories can be seen in Fig.~\ref{fig:MetProfiles}, and the results of this analysis are presented in Section \ref{sect:Results}. We chose to use these definitions, with dip and rise features separated out into extended or central based on the position of $R\sub{br}$, so that we could explore whether the galaxies best-fit with a central dip shared similar properties to those in the extended dip category, potentially suggesting some form of progression. It must be noted that, while we have chosen to name our categories as a `dip' or `rise', there are galaxies for which the fitted profile could be equally interpreted as a central flattening. Similarly, for the `extended dip / rise', there are instances where this could equally be described as an outer flattening. 

We note that \cite{Belfiore2017} caution against drawing physical conclusions from metallicity gradients within $\sim{}0.5R\sub{e}$ in MaNGA, due to the beam-smearing effects. We discuss the potential of beam-smearing effects on our results in Section \ref{sect:spatialresolution}.

\subsection{Ionisation Parameter Maps}
\label{subsect:IonParamMaps}

The anti-correlation between $\log(U)$ and metallicity is well-known (see Section \ref{sect:logU}), indeed it forms the basis of the O3N2 diagnostic \citep{Alloin1979, Kewley2002}. If the observed central metallicity dips were to be simply an artefact of the diagnostic used, one way this could arise is if the relationship between $\log(U)$ and 12+log(O/H) differed at the centres of galaxies compared to the relation present in samples used to calibrate strong line metallicity diagnostics. To investigate this possibility further, we produced $\log(U)$ maps for all galaxies in our sample, which we could then compare to our metallicity maps.

Determining the ionisation parameter usually requires comparing the strength of emission lines from two different ionisation levels of the same element. Ratios of the \oiii\ and \oii\ lines are widely used, as they have a large difference in ionisation potential, and therefore constrain the ionisation parameter well \citep[e.g.][]{Kewley2002}. However, this flux ratio is also known to have a strong dependence on the metallicity, making sulphur-based $\log(U)$ diagnostics preferable for our analysis, as they have been shown to have a much weaker metallicity dependence \citep{Kewley2002, Dors2011, Morisset2016}.

We chose to use the sulphur-based ionisation parameter diagnostic presented in \cite{Mingozzi2020}, as this diagnostic uses the \siii$\lambda\lambda$9070, 9531 lines suggested by \cite{Kewley2002} to provide the best measure of the ionisation parameter, and is calibrated against MaNGA data. Photoionisation models have been found to over-predict the strength of the \siii\ lines \citep{Mingozzi2020}, which \cite{Kewley2002} suggested to cause an under-estimation of the ionisation parameter compared to oxygen-based diagnostics. This effect is accounted for in the re-calibration presented by \cite{Mingozzi2020}. Using the sulphur-based diagnostic presented in \cite{Dors2011} was found to return consistent results.  

We produced flux maps for the \siii$\lambda\lambda$9070, 9531 lines following a similar process to that used within the DAP \citep{Belfiore2019}. For each galaxy, LOGCUBE files are available from the MaNGA Data Reduction Pipeline \citep[DRP;][]{Law2016}. The LOGCUBEs are 3-dimensional data cubes (2 spatial dimensions and one wavelength), with spectra covering the wavelength range 3600--10000 Å corresponding to each spatial pixel, and we used these to produce maps of the \siii$\lambda\lambda$9070, 9531 lines.

The Model LOGCUBE files provided as part of the DAP contain information on the required Galactic reddening correction \citep{Belfiore2017}, which we used to correct the spectra within the LOGCUBE. The stellar continuum models also provided within the Model LOGCUBE files do not extend beyond 7400 \AA. However in the wavelength region around the \siii$\lambda\lambda$9070, 9531 lines there are no prominent stellar absorption lines, so for our purposes of isolating the gas-phase emission, it was sufficient to approximate the continuum as flat, as done by \cite{Mingozzi2020}. The LOGCUBE files returned by the DRP are already sky background subtracted, and at wavelengths longward of $\sim 8500$\AA, where there is bright telluric line emission, the residuals from the sky background procedure are on the order of $10-20$\% \citep{Law2016}. The background residuals are propagated into the LOGCUBE error spectra, which we use to determine the corresponding uncertainty in our \siii$\lambda\lambda$9070, 9531 line flux measurements.

Having corrected the spectra for Galactic reddening, we then corrected them for host galaxy reddening, as described in Section~\ref{subsect:Diagnostics}, then fitted Gaussian profiles to the emission lines in the returned spectra, using the \texttt{fit\_lines} method from the specutils python package \citep{astropy:2018}. In-keeping with the methods used in the DAP, where the velocity of all lines were tied, we fixed the position of each line to the gas-phase velocity map supplied in the MAPS files. We also tied the widths of any doublets, and fixed the ratio of their amplitudes to known theoretical ratios. For the \siii$\lambda\lambda$9070, 9531 lines, we used a flux ratio of 2.47 \citep{Luridiana2015}. We then produced maps of $\log(U)$, using the sulphur-based diagnostic presented in \cite{Mingozzi2020}, and converting to the dimensionless ionisation parameter, $U$, using $U = q/c$ where both $q$ and $c$ have units of cm s$^{-1}$ (equation (\ref{eqn:IonParam})). Here, $S3S2$ refers to the flux ratio \siii$\lambda\lambda$9070, 9531 / \siii$\lambda\lambda$6717,32.

\begin{equation}
    \log(U) = \log \left( \dfrac{q}{c} \right) = \frac{\log(S3S2) + 5.70}{0.76} - \log(c)
    \label{eqn:IonParam}
\end{equation}

\subbib

\section{Variations in Metallicity Profiles}
\label{sect:Results}

\begin{figure*}
    \centering
    \includegraphics[width=1\linewidth]{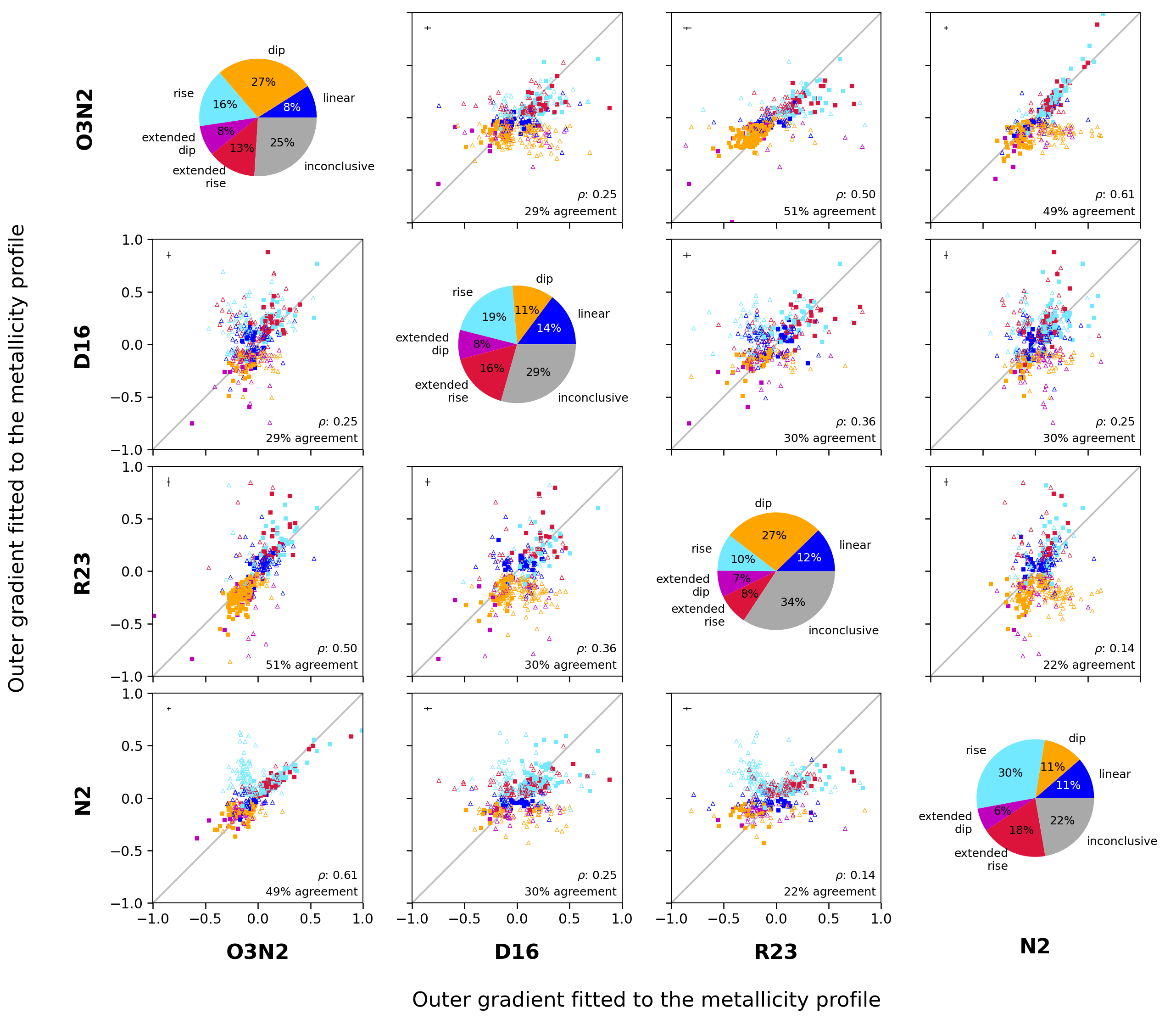}
    \caption{Along the central diagonal, pie charts show the percentage of galaxies within each metallicity profile category for each diagnostic. The gradient of the outer section ($R>R\sub{br}$) of the model fitted to each galaxy's metallicity profile (single gradient for linear) is represented in the scatter plots. The y-axis of each plot gives the gradient for the profile produced using the diagnostic given at the start of each row; the x-axis, the diagnostic given at the bottom of each column. The colours of the points represent the galaxy's category when the diagnostic on the y-axis is used, and the symbols represent whether the galaxy falls into the same category for both diagnostics considered (filled squares), or not (unfilled triangles). Finally, the Pearson's correlation coefficient ($\rho$), and the percentage agreement between the diagnostics (excluding any galaxies categorised as inconclusive for either diagnostic) are given in the bottom right-hand corner of each plot. The error bars were generally smaller than the data points, so average values are indicated in the top-left corner of each plot. }
    \label{fig:ComparingGradsouter}
\end{figure*}

\begin{figure*}
    \centering
    \includegraphics[width=1\linewidth]{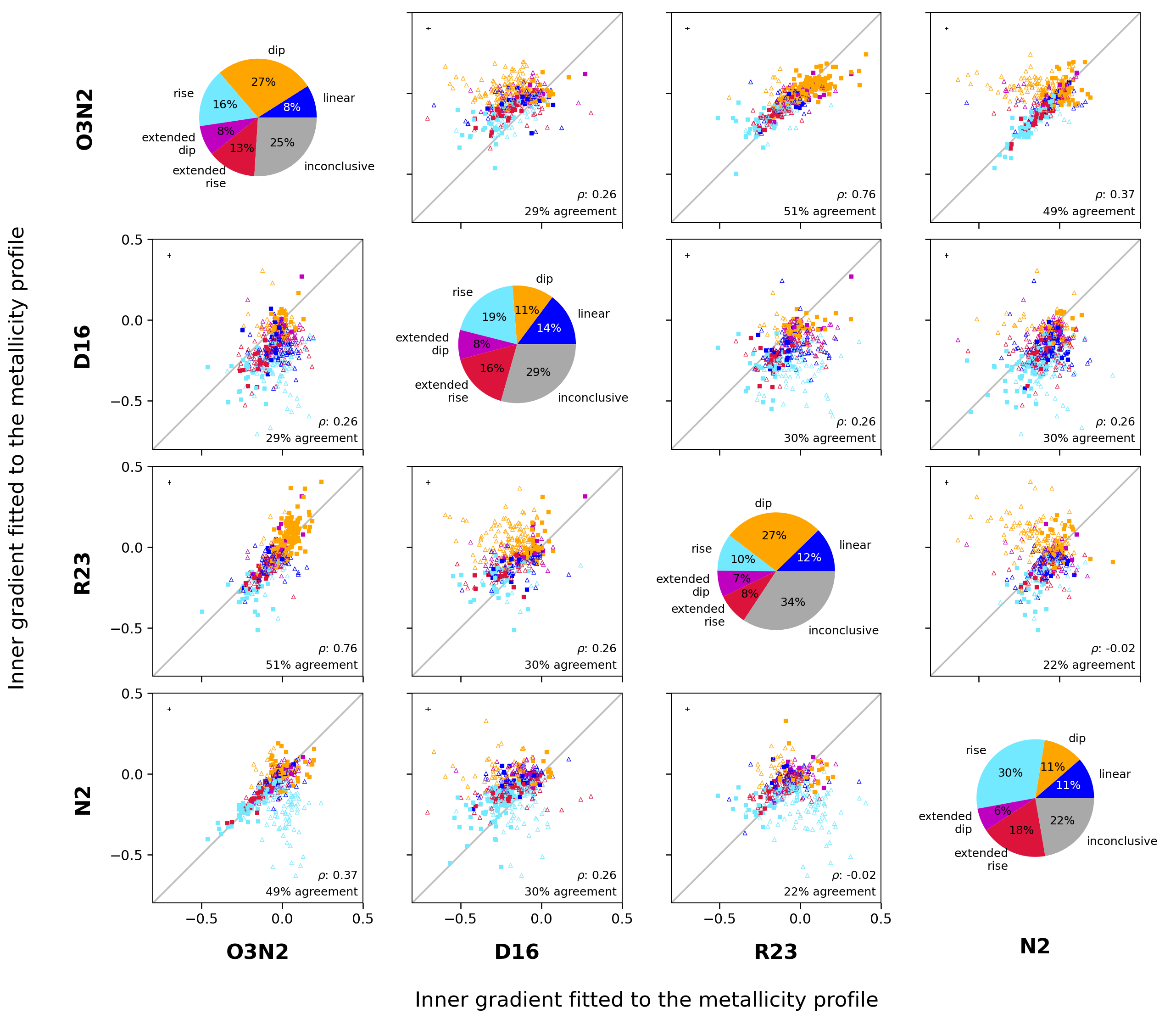}
    \caption{As for Fig.~\ref{fig:ComparingGradsouter}, but now comparing the best-fit $\alpha\sub{in}$ fitted to the radial metallicity profiles for different strong line diagnostics.}
    \label{fig:ComparingGradsinner}
\end{figure*}

To explore differences in the radial metallicity profiles produced when using different strong line diagnostics, we summarise our results from the best-fit models in Figs.~\ref{fig:ComparingGradsouter} and \ref{fig:ComparingGradsinner}. Along the central diagonal, pie charts indicate the percentage of galaxies falling within each metallicity gradient category for each of the four different metallicity diagnostics considered. Each scatter plot compares the best-fit metallicity gradient from two different strong-line diagnostics (indicated along the x- and y-axis) for each galaxy in our sample, where we only consider $\alpha\sub{out}$ (Fig.~\ref{fig:ComparingGradsouter}), or $\alpha\sub{in}$ (Fig.~\ref{fig:ComparingGradsinner}), in those cases where the metallicity radial profile was best-fit by a broken power law. Any galaxies which had inconclusive fits to the metallicity profile returned by either diagnostic are excluded from the scatter plots, and in order to zoom-in on the points within the plot, a small number ($\leq$5 per panel) of outlying points have been excluded. The uncertainties in the fitted gradients, taken from the covariance matrix returned from \texttt{curve\_fit}, were typically smaller than the data points, so are not shown. Instead, average uncertainties are indicated in the top-left of each plot. The points are coloured by the categorisation of the galaxy when considering the metallicity profile produced using the diagnostic on the y-axis, with filled squares indicating that both diagnostics agree on the categorisation of the metallicity profile, and unfilled triangles indicating that the two diagnostics disagree. The percentage agreement on profile classification between the two diagnostics (excluding any galaxies categorised as inconclusive for either diagnostic) is shown in the bottom right of each plot. We found that if we combined the categories of extended and central dips, and similarly for the rise categories, leaving us with only 3 different categories of galaxy metallicity profiles (linear, combined dips and combined rises), the agreement between pairs of diagnostics increased by $\sim$15\%.

As can be seen in these figures, we find that using different metallicity diagnostics causes large variations both in the fitted gradients and the categorisation of the galaxies, implying that the different diagnostics introduce different biases which in turn can affect the returned profiles. Galaxies for which both diagnostics agree on the categorisation (filled squares) appear to generally lie closer to the 1:1 line (grey line) than those for which the diagnostics disagree. This implies that the shapes of the returned metallicity profiles are indeed similar when two diagnostics agree on the categorisation of the profile. It is also apparent in these figures, that the gradients for galaxies categorised as having an extended dip appear to occupy a similar region to those having a central dip (magenta and orange data points, respectively), and similarly for the galaxies categorised as an extended or central rise (red and light blue respectively).

We also observe that around twice as many galaxies are categorised as having central dips when the O3N2 diagnostic is used, compared to the D16 diagnostic, where most of these galaxies are instead categorised as having linear or central rise profiles. Similarly, when the N2 diagnostic was used, around twice as many galaxies are categorised as having a central rise compared to the other diagnostics. In both of these cases, the galaxies exhibiting a central dip in the N2 or D16 profiles are not simply a subset of those with a central dip in the O3N2 or R23 profiles. This observed dependence of the profile shape on the diagnostic used is somewhat reflected in \cite{Yates2021}, where it was found that diagnostics relying on the use of the \oiii{} lines favoured profiles with a flattening in the central regions for higher mass galaxies, and diagnostics which did not rely on the use of these lines were more likely to favour a continued increase towards the centre. 

In both Fig.~\ref{fig:ComparingGradsouter} and Fig.~\ref{fig:ComparingGradsinner} we find that the D16 diagnostic seems to show a low level of agreement with all 3 of the other diagnostics ($\sim$30\%), as well as a large amount of scatter in the fitted gradients, as indicated by the Pearson correlation coefficient ($\rho$) given in the bottom-right corner of each scatter plot. This would appear to suggest that the shapes of the metallicity profiles as measured by the D16 diagnostic are the most different to those returned by the other diagnostics.

The O3N2 diagnostic, on the other hand, shows the highest levels of agreement with the other diagnostics, with $\sim$50\% agreement with the R23 and N2 diagnostics. These pairs of diagnostics also show the highest correlation in the fitted gradients for both the outer ($\rho$=0.50 and 0.61, respectively) and inner ($\rho$=0.76 and 0.37, respectively) fitted gradients, with the pie charts showing the  O3N2 and R23 diagnostics to also have a similar number of galaxies falling within each category. For the galaxies categorised as having a central dip in the O3N2 profile, 58\% of these were also categorised as having a dip by one other diagnostic, 20\% by two other diagnostics, and 2.4\% showed agreement by all 4 diagnostics. A higher level of agreement could be expected in the case of the O3N2 and N2 diagnostics, given that these diagnostics were calibrated on the same sample of galaxies, but instead we find that almost twice as many galaxies are categorised as having a central rise when using the N2 diagnostic, and half as many as having a central dip. 

Considering the subplots comparing the N2 or D16 diagnostics to the O3N2 and R23, a small cloud of points can be seen, where galaxies have been fitted with a central rise in the N2 or D16 diagnostics, but a central dip in the other two diagnostics. Further investigation into the possible causes of this is out of the scope of this work, but may be related to a change in relation between the N/O and O/H abundances \citep[e.g.][]{Schaefer2020}.

We found very good agreement between results using the \cite{Pettini2004} O3N2 diagnostic and the \cite{Curti2017} re-calibration presented here, both in terms of the categorisation of the galaxies, and the fitted gradients. This implies that the diagnostics presented by \cite{Curti2017}, which were calibrated against stacked unresolved data, are equally applicable to our spatially-resolved observations.

\subsection{Spatial Resolution Effects}
\label{sect:spatialresolution}

\begin{figure}
    \centering
    \includegraphics[width=0.9\linewidth]{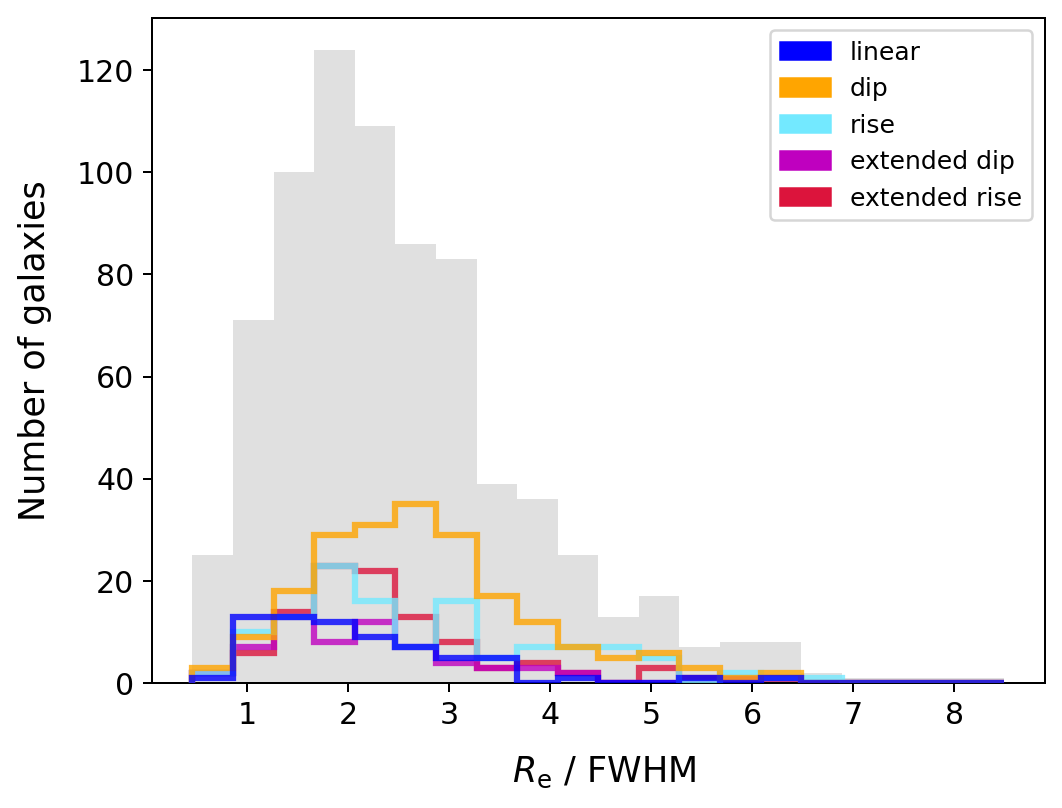}
    \caption{The distribution of the number of spatial resolution elements per $R\sub{e}$ is shown, with the sample of 758 galaxies shown as the light grey histogram, and galaxies falling within each O3N2 metallicity profile category shown in the various colours.}
    \label{fig:spatialres}
\end{figure}

There have been suggestions that beam-smearing effects in MaNGA lead to issues with fitted metallicity gradients in the central regions of galaxies. For example, \cite{Belfiore2017} found that including spaxels $\leq$0.5 $R\sub{e}$ caused a systematic flattening when fitting linear gradients. To explore whether this effect could be responsible for the observed central dips, we checked for any dependence with spatial resolution for galaxies categorised by their O3N2 metallicity profile. In Fig.~\ref{fig:spatialres} we compared the number of spatial elements per $R\sub{e}$ for galaxies within each category, calculated as the $R\sub{e}$ for each galaxy divided by the median FWHM of 2.54 arcsec \citep{Law2016}. If the galaxies with central dips were to have significantly lower values of $R\sub{e}$ / FWHM, then this could suggest that the lower metallicity values observed in the centre could be produced by beam-smearing effects \citep[see fig. 2 of][for example]{Belfiore2017}. We find no evidence of this, suggesting that while the beam-smearing could cause inaccuracies in the measured inner gradients, the lack of spatial resolution at small radii is not the sole cause of the dips. The fact that different metallicity diagnostics return a different prevalence of central-dip galaxies also suggests that beam smearing is not the dominant factor.

\subsection{Average Fitted Profiles}
\label{subsect:AvgFittedProfiles}

\begin{figure*}
    \centering
    \includegraphics[width=0.8\textwidth]{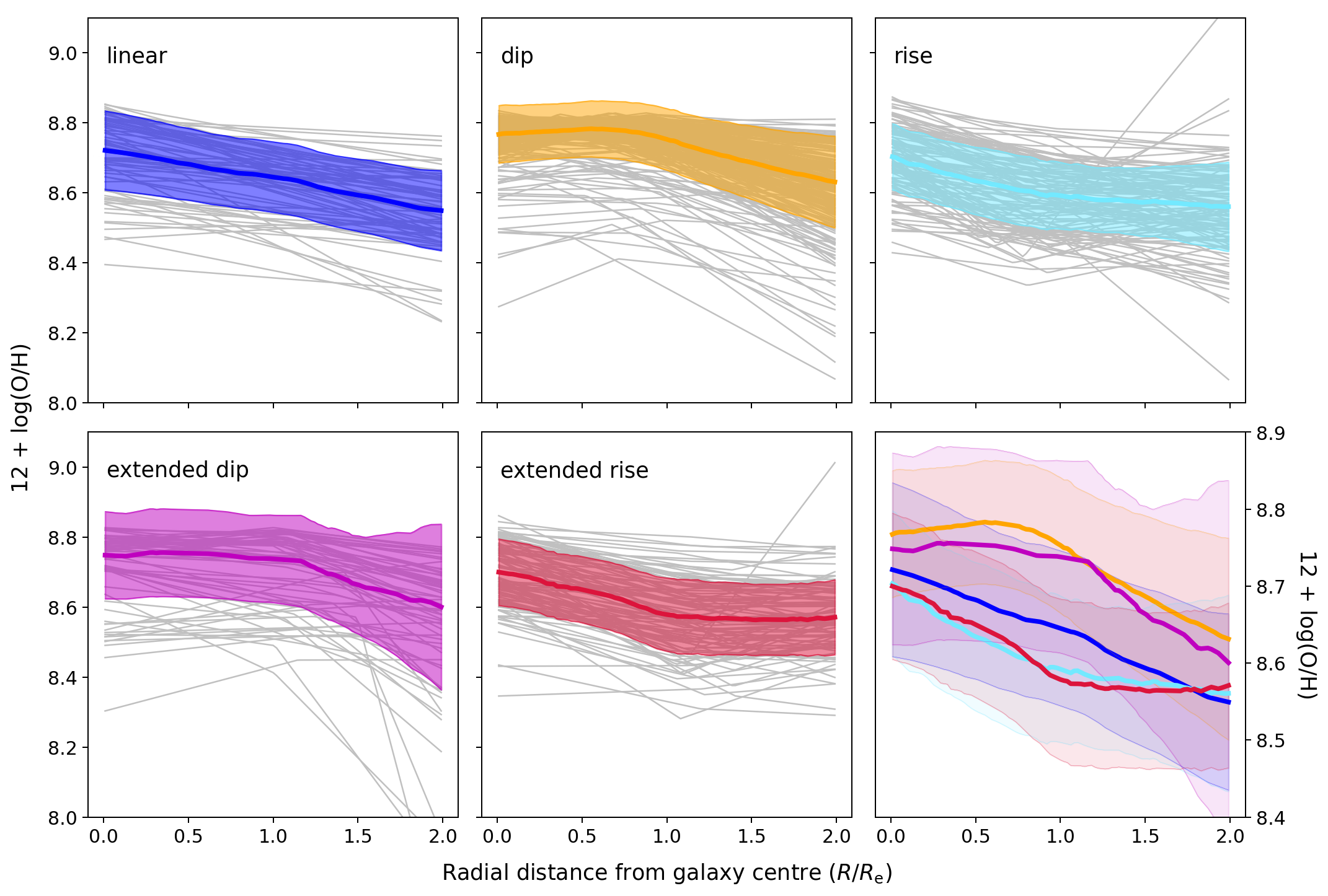}
    \caption{The models fitted to the O3N2 metallicity profiles for individual galaxies can be seen as the grey lines in each subplot, with the coloured lines showing the median profile for each category. The shaded region represents the rms of the residuals, to give a measure of the scatter from the individual fitted profiles. The bottom-right plot shows a comparison between the median profiles. Note that the scale on the y axis is slightly different here.}
    \label{fig:avgMet}
\end{figure*}
 
Each galaxy's fitted metallicity profile (when using the O3N2 diagnostic) is shown as grey lines in Fig.~\ref{fig:avgMet}. For each metallicity profile category, we produced median-averaged metallicity profiles from these best-fit models as a function of normalized (i.e., $R/R\sub{e}$) galactocentric radius, taking the median value within bins of width 0.1 $R\sub{e}$. The shaded regions represent the rms of the residuals, to give a measure of the scatter in the individual profiles.

From Fig.~\ref{fig:avgMet}, it can be seen that there is a range of radial profile shapes within each category, and that for galaxies categorised as having a central dip, the average metallicity is relatively constant out to 1$R\sub{e}$. The shift towards higher absolute metallicity values for galaxies categorised as having a central or extended dip is likely due to the higher global stellar mass of galaxies within these categories, as discussed in Section \ref{subsect:GlobalMassSFR}. The extended and central dip categories appear to return similar average profiles, as do the extended and central rises.

\subbib

\section{The Effects of Ionisation Parameter}
\label{sect:logU}

Before exploring possible physical causes for the observed central dips, we first investigate whether these features could be an artefact of the strong line diagnostic used, rather than representing a genuine decrease in the metallicity in these regions. Two of the main properties of the nebular gas which can affect the metallicity measured with strong line diagnostics are the ionisation parameter, and the N/O ratio. The effect of the N/O-O/H relationship on derived metallicity gradients has been explored by \cite{Schaefer2020}. Their fig.~1 indicates that the outer regions of more massive galaxies are offset in the N/O-O/H relation such that they have larger N/O values for the same given oxygen abundance. For metallicity diagnostics that assume a fixed N/O-O/H relation, such an offset in the outskirts of more massive galaxies could artificially introduce central rises into our profiles by flattening the outer regions, but this cannot be used to explain artificially introducing a central dip. We therefore instead explore how differences in the $\log(U)$-log(O/H) relationship could impact our measured gradients.

Systematic differences in the absolute metallicity values returned by differently-calibrated strong line diagnostics are well known \citep{Kewley2008}, and while joint calibrations such as those presented by \cite{Curti2017} reduce the systematic differences between diagnostics from $\sim$0.4-0.6 dex to $\sim$0.05 dex, we still observe differences in the gradients fitted to the N2 and O3N2 profiles in Figs.~\ref{fig:ComparingGradsouter} and \ref{fig:ComparingGradsinner}, implying that on spatially resolved scales, the consistency between diagnostics breaks down. If global or stacked spectra are used to calibrate strong line diagnostics, this can introduces biases, as these spectra tend to be luminosity-weighted, and therefore biased towards emission from high-temperature gas. If the conditions within the centres of galaxies are not well-matched to these conditions, then the diagnostics may be unreliable when used in these regions. 

There is a known anti-correlation between $\log(U)$ and the metallicity \citep[although, see][]{2021arXiv211000612J}, which has been hypothesised to arise from high metallicity gas having a higher opacity, and therefore acting to absorb more of the ionising photons \citep{Kewley2019}. Most strong line diagnostics rely on a fixed empirical or assumed relation between $\log(U)$ and 12+log(O/H). For example, this anti-correlation forms the basis of the O3N2 diagnostic, which relies on the \oiii\ and \nii\ lines, with O$^{++}$ and N$^+$ having large differences in the ionisation potential \citep{Alloin1979}. The \nii{} line is emitted predominantly from the low-excitation zone within H\textsc{ii} regions \citep{Kewley2002}, meaning the N2 ratio is strongly dependent on $U$. The R23 diagnostic also depends strongly on the ionisation parameter, as it uses both the \oii\ and \oiii\ lines, which have large differences in ionisation potential, and hence on the source's ability to ionise the surrounding gas \citep{Kewley2002}.

The O3N2 diagnostic has been found empirically to act as a good tracer of the metallicity, however if the photoionisation models presented by \cite{Kewley2002} reflect the full range of physical conditions observed within galaxies, we could expect to observe a larger amount of scatter in the measured metallicity values returned from the calibration samples than is often observed. This therefore raises the question of whether selection effects in the calibration samples could be removing some of this scatter.

Issues with the O3N2 diagnostic have been found when it is used on sub-HII region scales \citep[e.g.][]{Kruhler2017, Mao2018}, with measurements using this diagnostic returning metallicity values in the central region of observed HII regions $\sim$0.2-0.3 dex lower than that at the outskirts. This is in contrast to theoretical expectations of higher levels of metallicity in the centres of HII regions caused by hot young stars acting to ionise and enrich the surrounding gas. Both \cite{Kruhler2017} and \cite{Mao2018} therefore suggested that these measurements reflected inaccuracies in the diagnostic when used on this scale, possibly caused by changing ionisation conditions, rather than genuine decreases in the metallicity.

Previous works in which central metallicity dips were observed have mainly focused on results from using the O3N2 diagnostic \citep[e.g.][]{Sanchez2012, Sanchez2014, Sanchez-Menguiano2018}. The hypothesis that the dips in metallicity could be caused by the ionisation conditions present in the centre of the galaxy being somehow different to the range of conditions present in calibration samples therefore warrants careful consideration. 

We explored the effect of ionisation parameter on the returned metallicity values through two different approaches. Firstly, we binned spaxel spectra by their physical properties (\ie{}by $U$ and metallicity), and stacked the spectra in each bin. This allows us to compare the metallicity values returned by various strong line diagnostics to that returned by $T\sub{e}$ methods, to see whether the returned values could be shown to deviate significantly under certain physical conditions. Secondly, we focused on exploring whether the relationship between $\log(U)$ and 12+log(O/H) could be seen to differ significantly in the central regions of those galaxies categorised as having a central dip.

\subsection{Strong Line and $T\sub{e}$-based Metallicity Diagnostics}
\label{sect:StackedSpectra}

To investigate the effect of ionisation parameter on the various metallicity diagnostics considered in this paper, we use stacked spectra to compare the metallicity values obtained from various strong line diagnostics to those returned by $T\sub{e}$-based methods. The \oiiiauroral line is too faint to detect within individual spaxels, so we used stacked spectra in order to obtain sufficient S/N. Rather than using global spectra from individual galaxies, we take advantage of the spatially-resolved IFU data to stack spectra from regions of similar physical conditions within our sample of galaxies. When producing the stacked spectra, the physical conditions were defined by the inferred strong line metallicity and the ionisation parameter. It must be noted, however, that the $\sim$kiloparsec spatial resolution of MaNGA means that each spaxel comprises emission from more than just a single HII region, meaning that there may still be some effect of averaging over different conditions.

We chose to use the \cite{Curti2017} re-calibration of the N2 diagnostic to determine the metallicity of the spaxels when producing the stacks, as the relationship between the line ratio and metallicity is approximately linear \citep{Pettini2004}, so it does not have the complications of the double-branched nature of R23. The key goal of this binning procedure is to group spaxels of similar metallicity and $U$, therefore, the absolute value of metallicity in each bin (\ie{}the degree of accuracy of the metallicity diagnostic used) is less important here. The sulphur-based $\log(U)$ diagnostic presented in \cite{Mingozzi2020} was chosen to determine the ionisation parameter, for the reasons discussed in Section~\ref{subsect:IonParamMaps}. 

To select the sample of spaxels to be used to produce our stacked spectra, we first removed any spaxels which fell outside of the SF region of the BPT diagram, and any spaxels with S/N $<$ 5 in any of the lines used within the metallicity and $\log(U)$ diagnostics. We applied a lower limit on the H$\alpha$EW of the spaxels considered, removing any spaxels with H$\alpha$EW $<$ 50\AA, as used by \cite{Yates2020}. This was chosen as it was found to improve the measured S/N of the \oiiiauroral line in the corresponding stacked spectra. This left us with $\sim$46,500 spaxels, collected from all 758 galaxies.

\begin{figure}
\centering
\includegraphics[width=1\columnwidth]{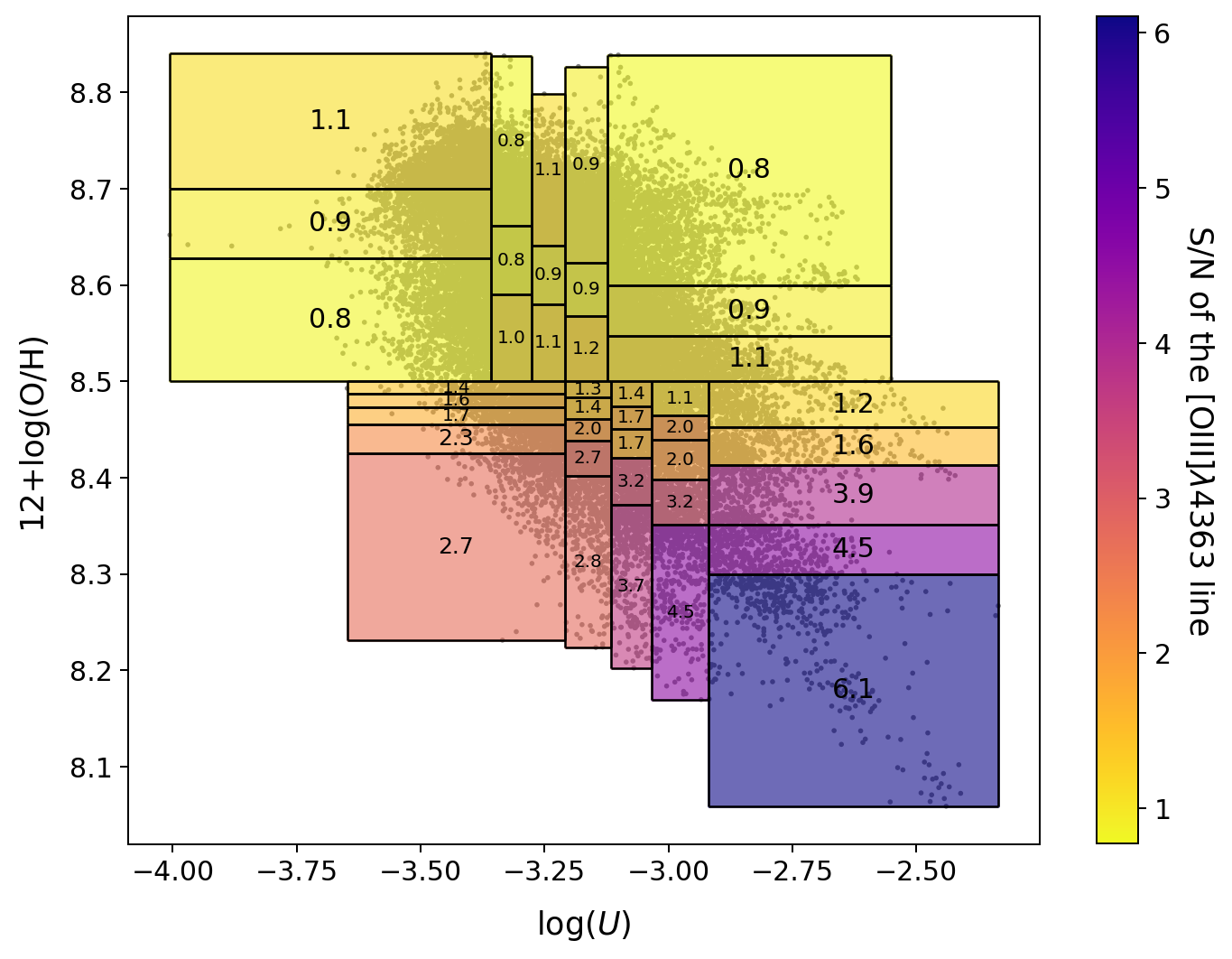}
\caption{The distribution of 12+log(O/H) and $\log(U)$ of the spaxels used to produce the stacked spectra can be seen as the grey points, with the bin limits overlaid as the black lines. The S/N of the \oiiiauroral line measured from the stacked spectrum resulting from stacking spaxels within each bin is shown as the text within each bin, and also by the colour.}
\label{fig:OIIISN}
\end{figure} 

Using the ionisation parameter and N2 metallicity maps, produced as described in Sections \ref{subsect:Diagnostics} and \ref{subsect:IonParamMaps}, we determined a binning scheme to group together spaxels encompassing gas under similar physical conditions. We tested several binning schemes, finding the scheme shown in Fig.~\ref{fig:OIIISN} to provide the optimal solution, with the best S/N of the \oiiiauroral line in the produced stacked spectra.

As the \oiiiauroral line becomes increasingly faint at high metallicities \citep{Kewley2019}, we found that dividing the spaxels into two halves, above and below 12+log(O/H) = 8.5, before binning them in terms of N2 metallicity and $\log(U)$ improved the S/N of the \oiiiauroral line in the final stacks. For the spaxels above 12+log(O/H) = 8.5, we split the spaxels firstly into 5 equally filled $\log(U)$ bins, and then split each column into 3 equally filled bins of metallicity, giving around 480 spaxels within each bin. For the spaxels below 12+log(O/H) = 8.5, the bins were determined in a similar fashion, with 5 bins of $\log(U)$, and 5 bins of metallicity. These lower-metallicity bins contained a greater number of spaxels, with around 2310 in each. The $\log(U)$ and metallicity distribution of the spaxels can be seen in Fig.~\ref{fig:OIIISN}, with the bins overlaid.

\begin{figure*}
    \centering
    \includegraphics[width=\linewidth]{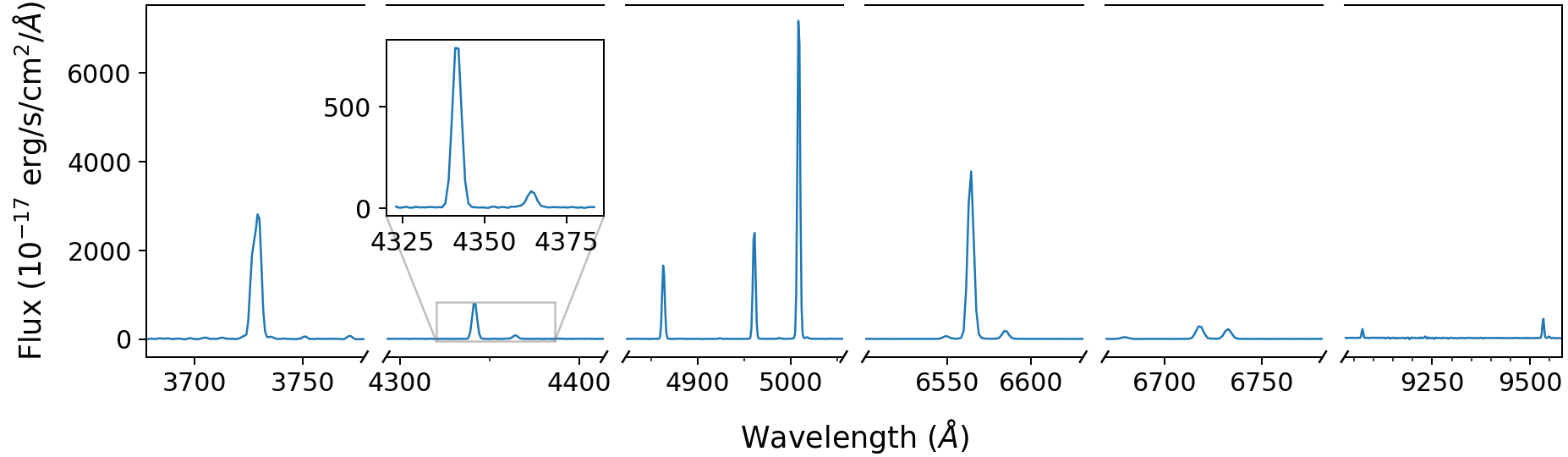}
    \caption{Example stacked spectrum, from spaxels falling within the bin $-2.92 < \log(U) < -2.33$, 8.05 $<$ 12+log(O/H) $<$ 8.30. The H$\gamma$ and \oiiiauroral lines are shown in the zoomed-in inset axis.}
    \label{fig:Example_stacked_spectrum}
\end{figure*}

\begin{figure*}
\centering
\includegraphics[width=1\textwidth]{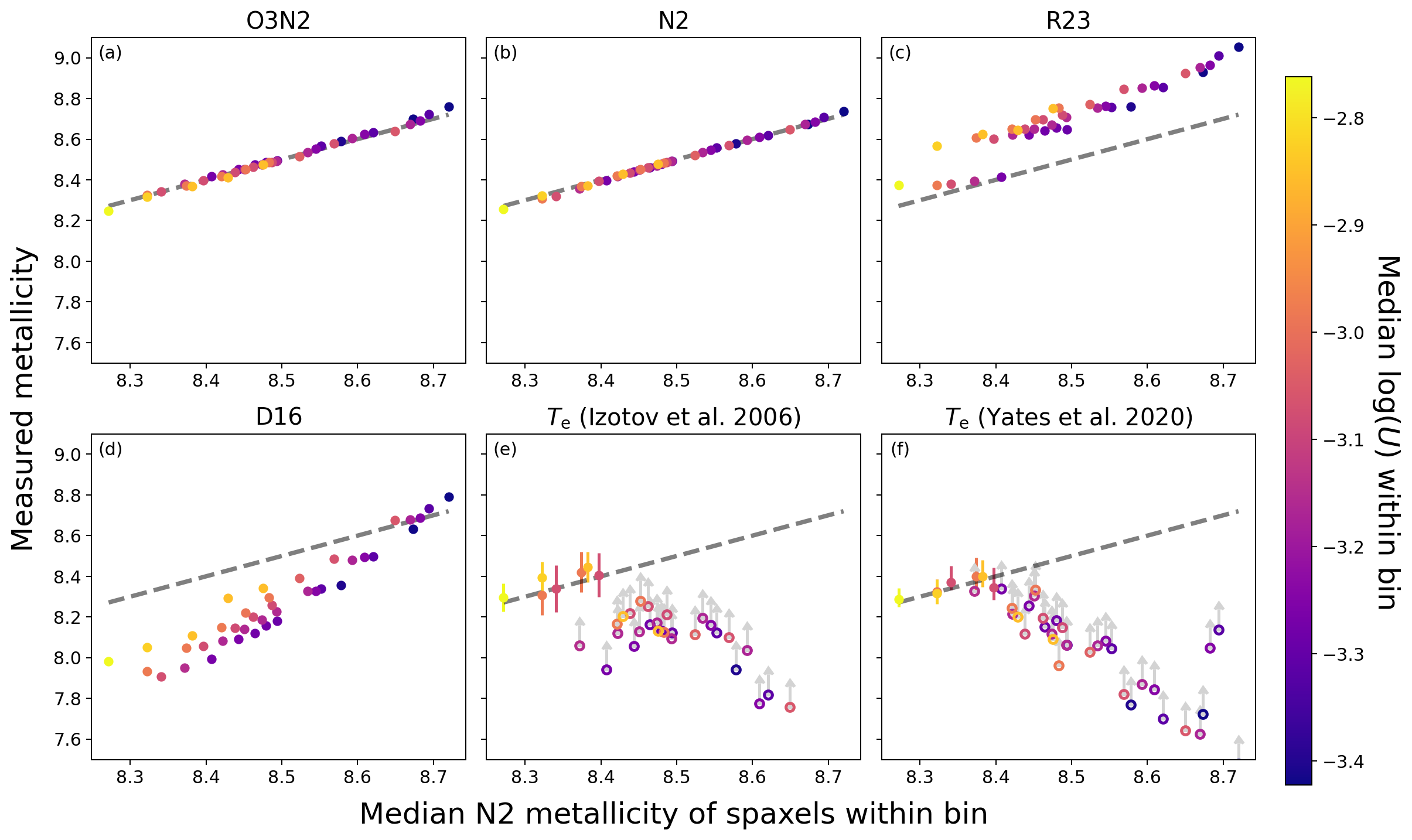}
\caption{The metallicity measured from the resulting stacked spectra is compared to the median value of the N2-based metallicity for spaxels within each bin. For the two $T\sub{e}$-based methods, measurements from spaxels with \oiiiauroral S/N $<$ 3 are given as $2\sigma$ lower limits, and shown as unfilled points. The points are coloured by the median $\log(U)$ of the spaxels within each bin.}
\label{fig:BinvMeasured}
\end{figure*}

Once the spaxels had been selected, and assigned to each bin of $\log(U)$ and metallicity, we used the data from the LOGCUBEs, which provide spectra covering the full range of 3600-10000\AA\ for each spaxel, to produce our stacked spectra. The LOGCUBE spectra were corrected for Galactic reddening, and the stellar component was removed as described in Section~\ref{subsect:IonParamMaps}. The spectra within each bin were then corrected for host galaxy attenuation, and spectra within each bin were stacked. An example stacked spectrum can be seen in Fig.~\ref{fig:Example_stacked_spectrum}. 

Gaussian profiles were then fitted to each emission line required for the strong line metallicity diagnostics, as well as for the $T\sub{e}$-based methods, following the process described in Section~\ref{subsect:IonParamMaps}. The width of the \oiiiauroral line was tied to that of the stronger \oiii$\lambda$4959,5007 lines when they were fitted, to reduce uncertainties in the fit to the fainter line. The measured S/N of the \oiiiauroral line for the stacked spectra is shown as the colours and also in overlaid text in Fig.~\ref{fig:OIIISN}. Despite investigating a number of different binning schemes, it was not possible to stack the spectra in such a way as to return more spectra with \oiiiauroral S/N $>$ 3. 

\cite{Curti2017} found that the \oiiiauroral line in their stacked spectra became increasingly contaminated by the \feii4360 line at higher metallicities, recommending that care must be taken to check for this contamination in spectra of high metallicity environments (12+log(O/H)$>$8.3). They showed that in the cases of \feii4360 contamination, the \feii4288 line was also clearly visible, therefore we visually checked each of our spectra for the \feii4288 line. We found no evidence for the presence of either of the \ion{Fe}{ii} lines in our stacked spectra, and so determined that our measurements of the \oiiiauroral line were not affected by this contamination. 

To determine the metallicity using the $T\sub{e}$-based method, we used two different diagnostics. We used the widely-established \cite{Izotov2006} diagnostic, and the recent \cite{Yates2020} diagnostic, which solves simultaneously for metallicity and the electron temperature of the O$^{+}$ gas (T$_{\oii}$), rather than assuming a fixed relationship between T$_{\oiii}$ and T$_{\oii}$. \cite{Yates2020} also used a diverse calibration sample, including both global spectra and spatially-resolved spectra for individual and composite HII regions, meaning their method should be equally applicable to systems with a wide range of physical sizes.

We measured the metallicity from each of the stacked spectra using the four strong line diagnostics detailed in Section \ref{subsect:Diagnostics}, as well as the two $T\sub{e}$-based methods. A comparison between the median metallicity of the spaxels within each bin using the N2 diagnostic, and the metallicity measured from the stacked spectrum, is presented in Fig.~\ref{fig:BinvMeasured}. For the $T\sub{e}$-based methods, where the \oiiiauroral line was detected with S/N $<$ 3, the measurement was converted to a $2\sigma$ lower limit, and are represented within Fig.~\ref{fig:BinvMeasured} as unfilled points.

All stacked spectra with the \oiiiauroral line detected with S/N $>$ 3 lie at metallicities 12+log(O/H) $\lesssim$ 8.4, and in this regime the N2 median metallicity of the stacked spaxels is in good agreement with both $T\sub{e}$-based metallicities (see panels e and f of Fig.~\ref{fig:BinvMeasured}). This result was found to be independent of the binning scheme used, but the binning scheme shown in Fig.~\ref{fig:OIIISN} maximised the number of bins with \oiiiauroral{} detected with S/N $> 3$. 

As expected, because they were calibrated to the same sample, the stacked metallicities based on the \cite{Curti2017} O3N2 diagnostic are also in excellent agreement with the median spaxel metallicities when using their N2 diagnostic across the sampled metallicity range. In the regime 12+log(O/H) $\lesssim$ 8.4, the metallicities returned by the O3N2 and N2 diagnostics are therefore in good agreement with the $T\sub{e}$-based measurements, suggesting that these strong line diagnostics are accurate for a range of $\log(U)$ values ($\sim{}-2.85$ to $-3.10$). Above this metallicity the $T\sub{e}$-based measurements are largely lower limits, but lie below the metallicities returned by the O3N2 and N2 strong line diagnostics, and are thus still consistent.

However, we do observe $\log(U)$-dependent discrepancies in both the R23 and D16 diagnostics relative to the N2 diagnostic, and therefore also relative to the $T\sub{e}$-based results at metallicities $<$ 8.4, despite the fact that both D16 and this calibration of R23 are designed to be independent of $\log(U)$. In Fig.~\ref{fig:BinvMeasured}, stacks from spaxels with lower $\log(U)$ values can be seen to return values of metallicity which are further away from the 1:1 line for these two strong line diagnostics. Consequently, the D16 diagnostic underestimates metallicities compared to the $T\sub{e}$-based metallicities by 0.2-0.3 dex, with this discrepancy increasing with decreasing $\log(U)$.

To ensure that our choice of metallicity diagnostic when deriving the stacked bins is not affecting our results, we repeated our stacking analysis using the D16-measured metallicity to bin the spaxels, and found similar results (see Appendix \ref{sect:Appendix_D16}), although the \oiiiauroral{} line could only be measured with S/N $\geq{}3$ for one bin in this case.

The shift in metallicity values returned by the R23 diagnostic above the 1:1 line likely reflects a known systematic discrepancy between diagnostics derived using photoionisation models and $T\sub{e}$-calibrated strong line methods \citep{Kewley2019}. The double-branched nature of this diagnostic can also be seen, with the upper and lower sequence of metallicity values returned from stacks with median metallicity values $<$ 8.4.

Based on these results, and those shown in Section~\ref{sect:Results}, we focus on results from the O3N2 diagnostic for our subsequent analysis.

\subsection{Relationship Between Metallicity and Ionisation Parameter}

\begin{figure*}
\centering
\includegraphics[width=1\textwidth]{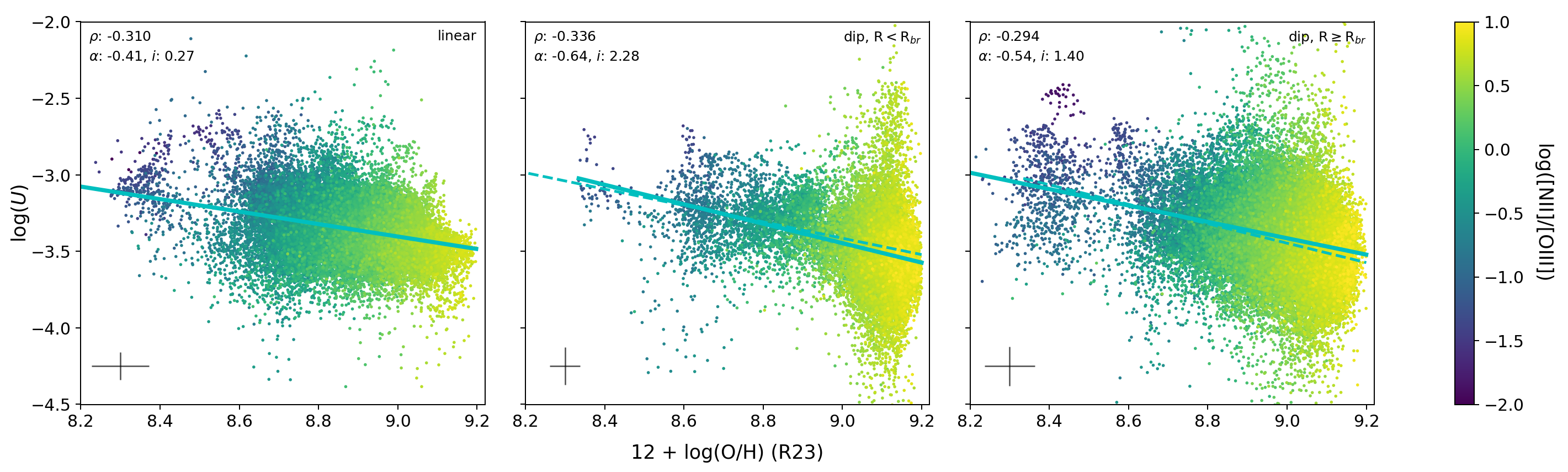}
\caption{In each plot, the relationship between $\log(U)$ and 12+log(O/H) (as measured by the R23 diagnostic) is shown, with the points coloured by the log(\nii$\lambda$6584/\oiii$\lambda$5007) flux ratio of each spaxel. The cyan line shows the best fit to the data; note that the fitted line here is not intended for use as a derived relation between $\log(U)$ and metallicity, instead it is used to aid comparison between the different subplots. The Pearson correlation coefficient ($\rho$) for $\log(U)$ and 12+log(O/H), and the gradient ($\alpha$) and intercept ($i$) for the fitted line, can be seen in the top-left of each plot. The average uncertainty on the points is shown in the bottom left. \textit{Left:} Spaxels are selected from galaxies categorised as linear in the O3N2 metallicity profiles, \textit{Central:} selecting only spaxels at $R < R\sub{br}$ for galaxies categorised as central dip, \textit{Right:} selecting only spaxels at $R \geq R\sub{br}$ for galaxies categorised as central dip. For the central and right plots, the solid line represents the best-fit to the data within that plot, and the dashed line represents the best-fit to the data on the adjacent plot, to aid comparison.}
\label{fig:UvOHSM20}
\end{figure*}

As we could find no clear evidence of the O3N2 and N2 diagnostics showing increasing discrepancies with the $T\sub{e}$-based methods under certain $\log(U)$ conditions, we instead explored whether there was any evidence for a differing relationship between $\log(U)$, 12+log(O/H), and log(\nii/\oiii) in the central and outer regions of galaxies exhibiting a central dip. The O3N2 and N2 strong line diagnostics rely on a single assumed relationship between these parameters, namely, the relationship that manifests in the particular calibration sample used. If there were to be evidence for a significantly different relationship in the central regions of our sample, this could cause measured metallicities to be unreliable. We emphasise that we make no attempt to derive a relationship between $\log(U)$ and 12+log(O/H). It is not possible to do so when using strong line metallicity diagnostics, as they themselves often include an assumed relationship between these two parameters \citep{Maiolino2019}.

In Fig.~\ref{fig:UvOHSM20}, we plot $\log(U)$ against R23-derived 12+log(O/H), coloured by log(\nii$\lambda$6584/\oiii$\lambda$5007), for the spaxels used to produce the radial metallicity profiles in Section~\ref{subsect:Diagnostics}. If the dips in metallicity were to be caused by a differing relationship between log($U$), log(O/H) and log(\nii/\oiii) in the central regions from that assumed in the strong line diagnostics, we would expect Fig.~\ref{fig:UvOHSM20} to show higher log($U$) values for a given value of log(\nii/\oiii) and 12+log(O/H) in the central regions (central plot), compared to the outer regions (right-hand plot) or linear galaxies (left-hand plot). For example, considering the results from the photoionisation model presented in \cite{Kewley2002}, fig.~8 shows that if the relationship between $\log(U)$ and 12+log(O/H) differed from that assumed by the diagnostic, this could lead to the metallicity being underestimated for a given flux ratio. There is a noticeable difference in the distributions, with a larger scatter in the $\log(U)$ values at high metallicity in the central regions, and reduced scatter at lower metallicities, compared to the spaxels at larger radii or within galaxies categorised as linear. However, we find no evidence for a change in relationship, from visually inspecting either the distributions or the fitted guide lines, suggesting that this cannot be used to explain the dips as an artefact of the strong line diagnostics. Plotting the relationships with the metallicity instead measured using D16 on the x-axis again shows no evidence for a change in relationship.

\subbib

\section{Dependence on Galaxy Properties}
\label{sect:GalaxyProperties}
\subsection{Global Properties}
\label{sect:Global}

As no clear $U$-dependent bias could be found to prove that the central dips are an artefact of the strong line diagnostic used, we now turn our attention to investigating whether the galaxies with central dips exhibit differences in any global properties compared to the rest of the sample. We use the O3N2 diagnostic to determine the categorisation of galaxy metallicity profiles in this section as it was found to be consistent with $T\sub{e}$-based metallicities for our stacked spaxel spectra (see Section \ref{sect:StackedSpectra}), and also has a reasonable level of agreement with the profile categorisations returned by the N2 and R23 diagnostics (49\% and 51\%, respectively), as shown in Section~\ref{sect:Results}. The R23 diagnostic would have the benefit of explicitly accounting for the $\log(U)$ dependence, however, it has the complication of being double-branched. The O3N2 diagnostic has a strong dependence on the ionisation parameter, therefore while it is not possible to assess the accuracy of the diagnostics using the results presented in Section~\ref{sect:Results}, the reasonable agreement between the O3N2 and R23 diagnostics despite the differences in $\log(U)$ dependence, could be suggestive that these diagnostics are tracing the metallicity well.

Our interpretations of the physical properties are therefore based upon the assumption that the O3N2 diagnostic is returning an accurate picture of the radial metallicity profiles. We revisit the impact of our choice of metallicity diagnostic on our results in Section~\ref{sect:Discussion}.

\subsubsection{Global Stellar Mass and Star Formation Rate}
\label{subsect:GlobalMassSFR}

\begin{figure*}
\centering
\includegraphics[width=1\linewidth]{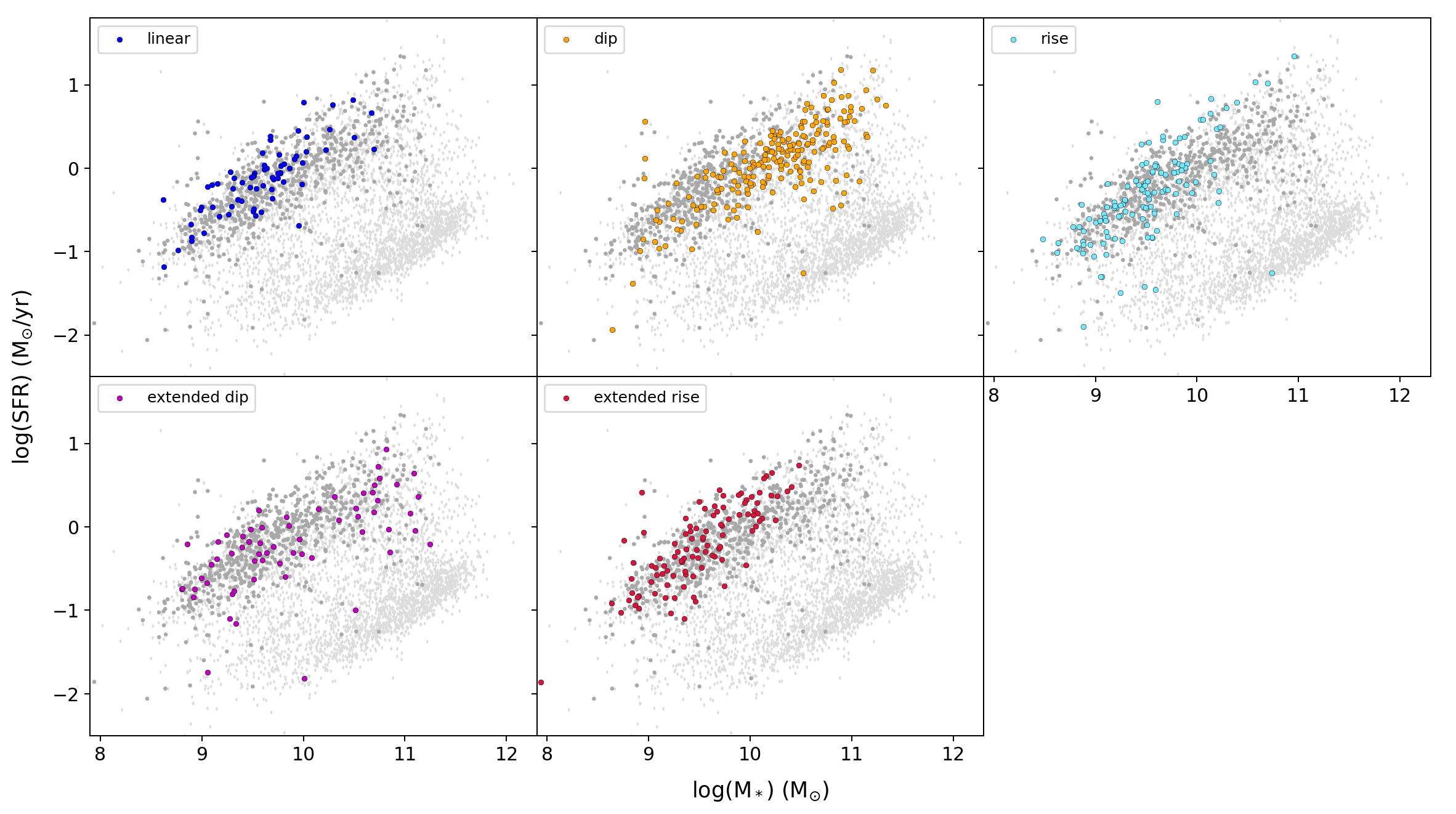}
\caption{The distribution of galaxies on the log(SFR) vs log(M$_*$) plane is shown, with the full parent sample of 4248 galaxies plotted as the light grey diamond symbols, and the sample of 758 galaxies shown as the darker grey points. The coloured points on each subplot represent the galaxies falling within each different O3N2 metallicity profile category. The galaxies categorised as having a dip are clearly shifted towards higher mass and lower SFR compared to the overall star-forming galaxy population, as also illustrated by Fig.~\ref{fig:global_logMstar_sSFR}}
\label{fig:MainSeq}
\end{figure*}

In Fig.~\ref{fig:MainSeq} it can be seen that the majority of our sample of galaxies lie on or above the main sequence, as is expected based on our sample selection criteria. Notably, a larger fraction of galaxies best-fit with dips appear to be located in the green valley, whereas galaxies with linear or rising proﬁles are comparatively more common among the upper half of the main sequence distribution.

\begin{figure*}
\centering
\begin{subfigure}[t]{.5\textwidth}
  \centering
  \includegraphics[width=.95\linewidth]{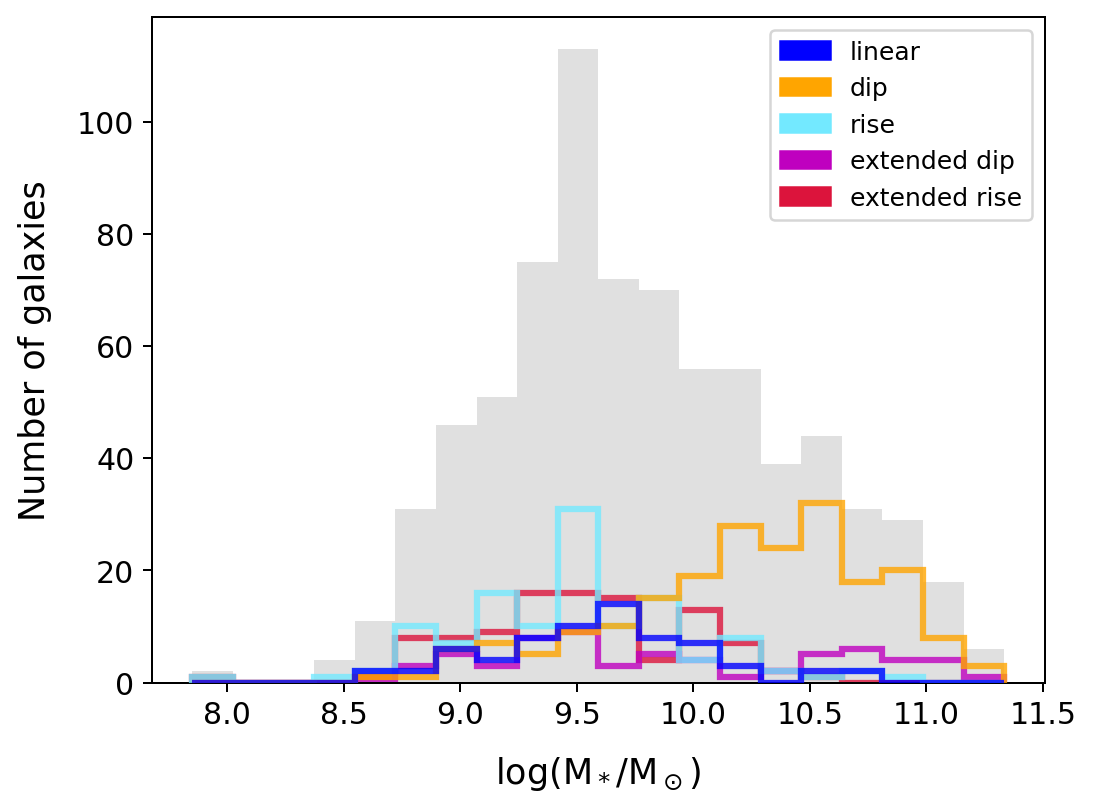}
\end{subfigure}%
\begin{subfigure}[t]{.5\textwidth}
  \centering
  \includegraphics[width=.95\linewidth]{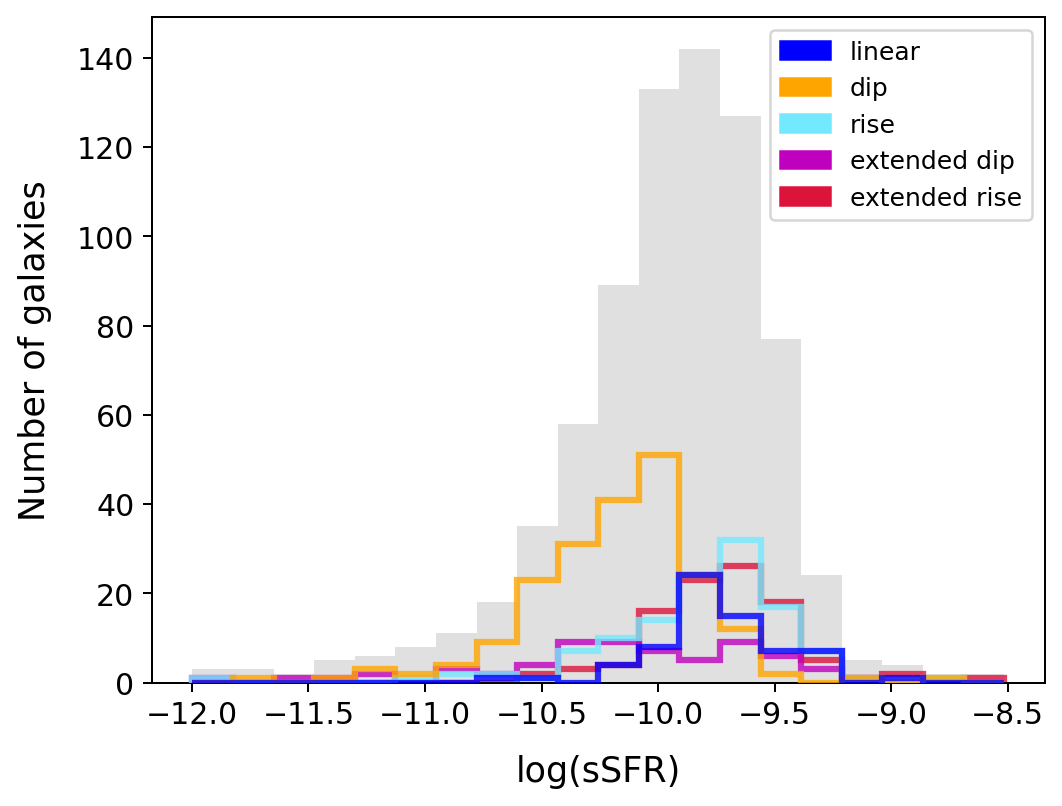}
\end{subfigure}
\caption{The sample of 758 galaxies is shown as the grey histogram, with the distribution of galaxies falling in each different O3N2 metallicity profile category shown in the various colours. \textit{Left:} The global stellar mass of the galaxies is shown, \textit{Right:} global sSFR values. A clear shift can be seen in the distributions of galaxies best fit by a dip towards higher stellar masses, and lower sSFR relative to the parent sample.}
\label{fig:global_logMstar_sSFR}
\end{figure*}

Several works have found that in median radial profiles for galaxies grouped by global stellar mass, the central dip becomes increasingly pronounced with increasing stellar mass \citep{Sanchez-Menguiano2016, Belfiore2017, Schaefer2019, Mingozzi2020}. As shown in Figs.~\ref{fig:MainSeq} and \ref{fig:global_logMstar_sSFR}, we also find that when comparing the sub-sample of galaxies exhibiting a central dip in the O3N2 metallicity profile to the parent sample of 758 galaxies shown in dark grey, there is a clear shift of galaxies with dips towards higher stellar masses. The global stellar masses reported in the MPA-JHU catalogue have been shown to be underestimated compared to more recent results for galaxies below $\logMm{}\sim{}10.0$ \citep{Blanton2011}, however this discrepancy is suggested to be on the scale of a few tenths dex, and therefore cannot account for the difference of $\sim$ 1 dex between the peak of the mass distribution for galaxies exhibiting a central dip, compared to galaxies within other categories.

Additionally, we find that galaxies with dips are also shifted towards lower sSFR, as seen in Figs.~\ref{fig:MainSeq} and \ref{fig:global_logMstar_sSFR}, which could suggest that quenching is occurring within the galaxy. It must be noted that as the sample of galaxies is chosen to include only galaxies exhibiting significant line emission and with star formation as the dominant ionisation method any signature of quenching will necessarily be subtle.

\subsubsection{Bulge to Total Ratio}

\begin{figure}
\centering
\includegraphics[width=1\linewidth]{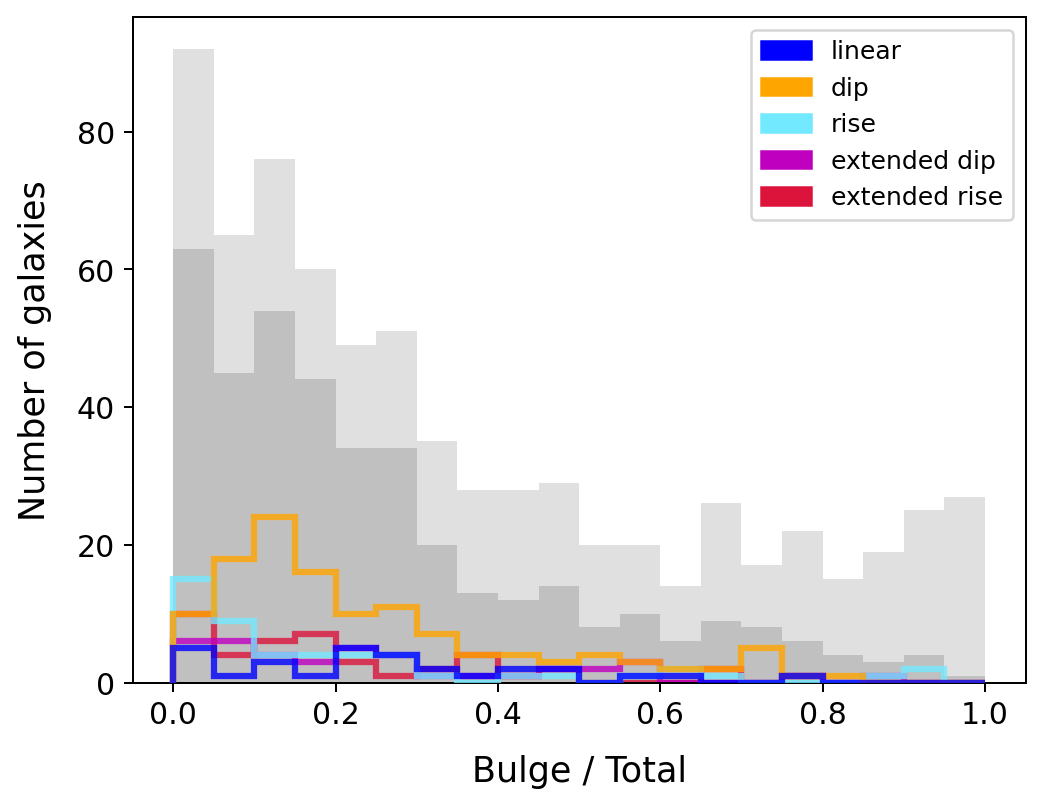}

\caption{The rest-optical $B/T$ ratio for the sample of 758 galaxies is shown in light grey, with galaxies having a bulge S\'ersic index $>$2, used to select out those with a classical bulge, shown in dark grey. Galaxies falling within each O3N2 metallicity profile category are shown in the various colours.}
\label{fig:global_BulgetoTotal}
\end{figure} 

To further investigate any possible link between the observed central dip and quenching, we investigated whether the galaxies exhibiting a central dip showed any difference in the bulge-to-total light ratio ($B/T$), compared to the rest of the sample. We used the measured $B/T$ values taken from the \cite{Simard2011} catalogue, which are quantiﬁed by 2D surface brightness proﬁle ﬁtting on the SDSS $r$-band images with a pure exponential disk, and a bulge component where the S\'ersic index was a free parameter. The distribution of $B/T$ values for galaxies categorised as having a central dip is visually slightly shifted towards higher values in Fig.~\ref{fig:global_BulgetoTotal}, although when comparing median values, the linear galaxies have a median of 0.24, compared to 0.18 for galaxies with a dip. The $B/T$ ratios therefore do not provide clear evidence of a link between central quenching and metallicity dips.

\subbib

\subsection{Spatially-Resolved Properties}
\label{sect:SpatiallyRes}

Given the possibility of inside-out quenching as the origin of the central dips, we searched for other signatures in the spatially-resolved physical properties of the galaxies. To do this, we plotted out average radial profiles for the H$\alpha$EW and D$_N$(4000) index, which act as powerful indicators of the star formation history over different periods within the galaxy \citep{Wang2017}, grouping the galaxies by their O3N2 metallicity profile category. The H$\alpha$EW was chosen as it acts as a tracer for the sSFR, with the H$\alpha$ flux acting to trace recent star formation, and the continuum level scaling with stellar mass \citep{Wang2020}.

The D$_N$(4000) break is caused by a number of absorption lines within a narrow wavelength region, with the absorption occurring within stellar atmospheres, and the break becoming increasingly pronounced with increasing stellar age. The amplitude of the break is therefore used to trace the age of stellar populations following recent star formation \citep{Balogh1999, Kauffmann2003}. We use the narrow band break, D$_N$(4000), as opposed to the wider band D(4000), as the wider band measurement is more sensitive to reddening effects \citep{Bruzual2003}. A map of D$_N$(4000) is provided in the DAP files, and uses the prescription presented in \cite{Balogh1999}.

\subsubsection{Average Radial Profiles}

\begin{figure*}
    \centering
    \includegraphics[width=0.8\textwidth]{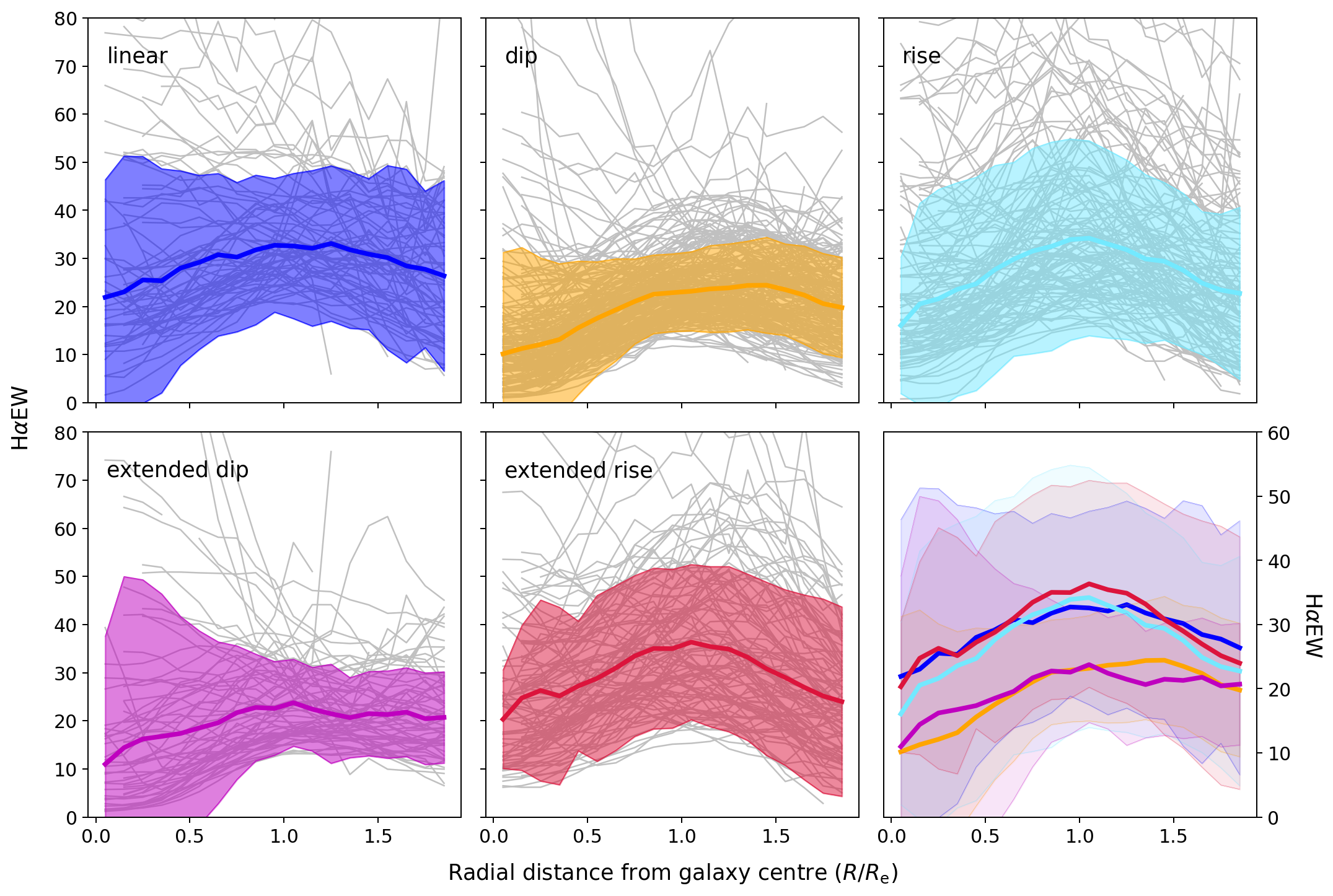}
    \caption{Average H$\alpha$EW profiles, with weighted mean profiles of individual galaxies shown as the grey lines, and the coloured lines representing the median of these profiles. The shaded region represents the rms of the residuals, and the bottom right subplot shows the median profiles for each different category overlaid for comparison - note that the scale on the y axis is slightly different for this plot. The H$\alpha$EW for galaxies with central and extended metallicity dips appear to be systematically lower than for the other galaxies, especially at $<1.5 R\sub{e}$.}
    \label{fig:avgHaEW}
\end{figure*}

\begin{figure*}
    \centering
    \includegraphics[width=0.8\textwidth]{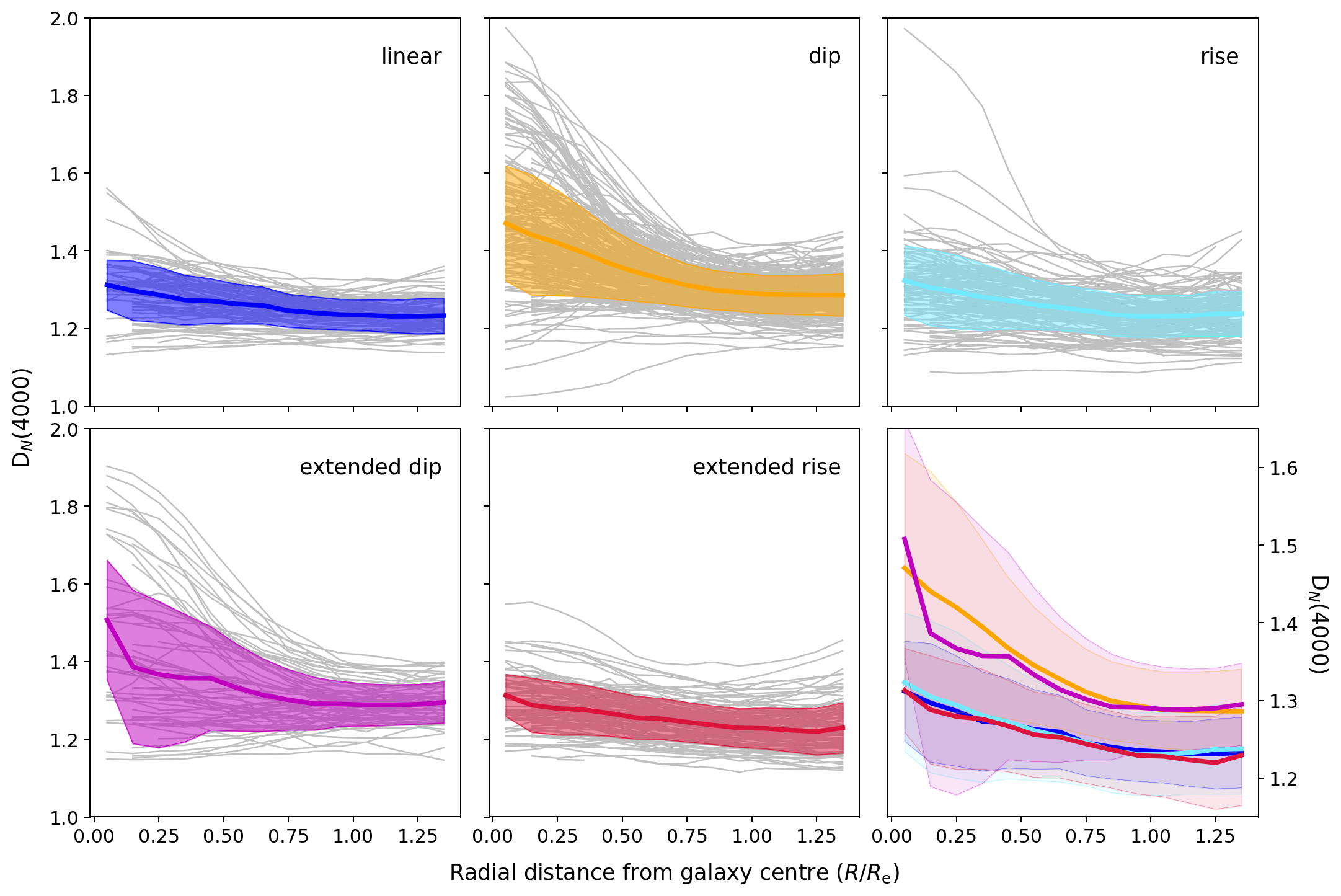}
    \caption{As for Fig.~\ref{fig:avgHaEW}, but showing the average D$_N$(4000) profiles, rather than the H$\alpha$EW. The D$_N$(4000) for galaxies with a central or extended metallicity dip appear to be higher in the central regions than for galaxies in other categories.}
    \label{fig:avgDN4000}
\end{figure*}

To produce azimuthally-averaged radial profiles within individual galaxies, we took the error-weighted mean within 0.1 $R\sub{e}$ width annuli, shown as the grey lines in Figs.~\ref{fig:avgHaEW} and \ref{fig:avgDN4000}. The median profiles, shown as the coloured lines, were then produced as described in Section \ref{subsect:AvgFittedProfiles}.

Comparing the median radial H$\alpha$EW profiles presented in Fig.~\ref{fig:avgHaEW} for galaxies categorised as having a dip or an extended dip to galaxies falling in the other categories, the H$\alpha$EW appears to be systematically shifted to lower normalisation, compared to galaxies falling within the other categories, especially below $\sim$1.5 $R\sub{e}$. As H$\alpha$EW is a tracer for the age of stellar populations, this could suggest a lack of recent star formation in the central regions leading to less H$\alpha$ flux being emitted, or alternatively a larger number of older stars increasing the continuum in these regions. Inflowing pristine gas has been suggested as a possible cause of the metallicity dips \citep{Sanchez2014}, however this could be expected to prompt star formation which would lead to higher values of H$\alpha$EW in the central regions, in contrast to the trends seen in Fig.~\ref{fig:avgHaEW}. Quenching in galaxies has been widely studied, and suggested to occur from the inside out in certain galaxies \citep[e.g.][]{Belfiore2017, Ellison2018}, potentially as a continuation of the inside-out growth \citep{Lian2017}. Although, we note that some simulations suggest SFR remains highest in the centres of galaxies, even during quenching \citep[e.g.][]{Henriques2020}.

When considering the average profiles of D$_N$(4000) in Fig.~\ref{fig:avgDN4000}, higher D$_N$(4000) values can be seen in the central regions of galaxies identified as having a central or extended metallicity dip. \cite{Belfiore2015} found young star forming regions to be associated with values of 1.2 $<$ D$_N$(4000) $<$ 1.4, and \cite{Wang2017} used D$_N$(4000) $> 1.6$ to define quenched spaxels / regions of the galaxies, dominated by older stellar populations. The higher D$_N$(4000) values observed in the central regions could therefore suggest that galaxies categorised as exhibiting a dip had relatively little central star formation over the past few hundred Myr to $\sim{}2$ Gyr ago. It would also appear that almost all of the galaxies with quenched central regions, according to the \cite{Wang2017} definition, fall within the central or extended dip categories in the O3N2 profiles, although by this definition not all galaxies with dips are quenched in the centre. 

To investigate whether there are signatures of a downturn in SFR within the past $\sim$Gyr, we compared average profiles of the SFR change parameter introduced by \cite{Wang2020}, finding no significant difference between different metallicity profile categories. This suggests that if quenching is occurring within these galaxies, the onset of quenching likely occurred more than $\sim$1 Gyr ago.

\subsection{Change in Slope vs Galaxy Properties}

Given the possible connection between the presence of dips in the O3N2 metallicity profiles and generally lower H$\alpha$EW and higher D$_N$(4000) values, we further investigated whether the strength of the dip ($\Delta\alpha$, defined as $\alpha_{in}$ - $\alpha_{out}$), showed any dependence with galaxy properties. When considering galaxy integrated properties, we found no correlation between the strength of the dip and with either the global stellar mass or star formation rate (rank correlation coefficient $\rho \sim$ 0.2 and 0.01, respectively).

Considering the spatially resolved properties, we produced average radial profiles of H$\alpha$EW and D$_N$(4000) for the galaxies exhibiting a central dip in the O3N2 metallicity profiles (see Appendix~\ref{Appendix:DeltaSlope}). The galaxies were split into 6 equally-spaced bins according to their $\Delta\alpha$, with median profiles produced for each bin. In Fig.~\ref{fig:Appendix_HaEWDN4000avg}, a tentative trend towards lower normalisation in H$\alpha$EW can be seen with increasing $\Delta\alpha$, although the highest $\Delta\alpha$ bins do not follow this same trend across all radii. Similarly, a tentative correlation can be seen between increasing strength of the dip and higher values of D$_N$(4000), although again the highest $\Delta\alpha$ bins deviate from this. These trends appear to be independent of the binning scheme chosen to group the galaxies by $\Delta\alpha$.

\subbib

\section{Discussion}
\label{sect:Discussion}

In the results presented in Section~\ref{sect:GalaxyProperties} we find evidence, when using the O3N2 diagnostic to categorise radial metallicity profiles, that there is a higher prevalence of central dips in galaxies with higher stellar masses and lower specific star formation rates. We additionally find that these galaxies exhibiting a central dip have lower H$\alpha$EW and higher central D$_N$(4000) values on average, and find tentative evidence for galaxies with stronger dip signatures having correspondingly lower H$\alpha$EW and higher D$_N$(4000) values, although this trend does not extend to the largest $\Delta\alpha$ bins. If the trends observed when using the O3N2 diagnostic accurately reflect radial trends in the metallicity, these results may imply that central quenching within the galaxies is linked to the presence of central metallicity dips.

By studying radial profiles of the SFR surface density, \cite{Ellison2018} showed that galaxies could be expected to both grow and quench from the inside-out, presenting several possible mechanisms for central quenching. Central quenching could act to cause these central metallicity dips either due to lower SFR in the central regions slowing down the enrichment of gas, or due to metal-rich outflows preferentially removing gas from the central regions of these galaxies. As our `dip' category includes both galaxies exhibiting a clear drop in the metallicity in the central regions, as well as those exhibiting a central flattening, a combination of various mechanisms could be acting. Reduced enrichment relative to that at greater radii could cause a flattening, and outflows of metal-rich gas may cause a reduction in the central metallicity.

Gaseous outflows have been observed to be generally centrally-concentrated in studies of MaNGA galaxies \citep{Avery2021}, despite the deeper gravitational potential in the central region of galaxies. Although these outflows may not exceed the escape velocity, they may still cause redistribution of enriched gas and thus alter metallicity profiles. We investigated the prevalence of gaseous outflows among the galaxies in our sample, as evidenced by broad velocity components to the strong rest-optical emission lines \citep[see][for details on methodology]{Avery2021}. Galaxies were defined as having an outflow if including a broad component when fitting the emission lines gave a statistically significantly better fit, with additional requirements placed on the broad-to-narrow flux ratio, and the velocity dispersion of the broad and narrow components \citep{Avery2021}. The subset of galaxies with detectable galactic winds in the ionised gas phase is low ($\sim 6\%$) among our sample of star-forming galaxies without clear signatures of AGN activity, and importantly spans almost all of the metallicity profile categories rather than showing a higher incidence rate for certain categories. Between the different categories, the incidence varies from 8\% for linear and central dip profiles to $\leq$5\% for the other categories.  Given Poisson uncertainties of typically $<$1\% on the respective incidence rates \citep{Gehrels1986}, we conclude that there is no significant evidence for outflow incidence being associated with particular metallicity profile types. 

The lack of evidence for galaxies with central dips in the O3N2 metallicity profiles having stronger outflows, notably larger \textit{B/T} ratios or evidence for quenching in SFR change parameter profiles, leaves any connection between dips in galaxy metallicity gradients and quenching far from substantiated. An alternative cause for the flattening of metallicity profiles in the central regions of galaxies could be the presence of bars, which cause increased mixing of gas \citep{Zurita2021}. To test this hypothesis, we used the visual morphological classifications from the SDSS and DESI images, taken from the MaNGA Value Added Catalogues. This catalogue uses the methods from \cite{VisualMorph}. Based on the Hubble classifications of our sample, there was no compelling evidence of galaxies with a dip in the metallicity profiles produced by any of the strong line diagnostics having a higher prevalence of bars. In fact, contrary to expectations of a central bar flattening the inner gradients, the highest prevalence of bars was consistently observed in the rise and extended rise galaxies when considering the categorisation according to all four strong line diagnostics. For the galaxies categorised with a central dip, 13--25\% of the galaxies had evidence of a strong bar for the four diagnostics, compared to 28--40\% and 25--30\% for those with a central and extended rise, respectively.

Given the differences that we found in the metallicity profiles of galaxies according to the metallicity diagnostic used, we cannot rule out that the prevalence for lower H$\alpha$EW and larger D$_N$(4000) in galaxies with central O3N2 metallicity dips is because these are the conditions under which the O3N2 diagnostic breaks down. All strong line diagnostics have benefits and drawbacks \citep[e.g.][]{Maiolino2019}, for example the R23 diagnostic does not require the use of a proxy element such as nitrogen, but is double-valued in nature, which we observed to cause issues with fitting gradients to our metallicity profiles. The N2 diagnostic does not have this same complication, but is known to saturate at high metallicities, which could cause dips to be artificially introduced if the inner regions reach saturation \citep[e.g.][]{Teimoorinia2021}. Given that large differences in the categorisation of the profiles were observed when different diagnostics were used (Figs. \ref{fig:ComparingGradsouter} \& \ref{fig:ComparingGradsinner}), we investigated how using the D16 diagnostic, instead of the O3N2, to categorise the galaxies affected the observed global and spatially-resolved properties. The D16 diagnostic showed the lowest agreement with O3N2 on the categorisation of galaxies, as well as having a large difference in the percentage of the sample exhibiting a central dip, with only 11\% of galaxies categorised as having a central dip when the D16 diagnostic was used, compared to 27\% for the O3N2 diagnostic. The galaxies with a central dip in the D16 profiles were also not simply a subset of those exhibiting a dip in the O3N2 profiles; in fact, only 36\% of the galaxies with a dip in the D16 profile also had a dip in the O3N2.

The global and spatially-resolved parameter distributions, using D16 to categorise the galaxies, are presented in Appendix \ref{Appendix:D16}. They display clear differences compared to the results presented in Section~\ref{sect:GalaxyProperties}, with galaxies exhibiting a central dip no longer being distinct in terms of their stellar masses and specific SFRs, or H$\alpha$EW and D$_N$(4000) profiles. Re-producing Fig.~\ref{fig:Appendix_HaEWDN4000avg}, for the galaxies categorised as having a dip in the D16 metallicity profile, there is no longer any clear trend between $\Delta\alpha$ and the normalisations of H$\alpha$EW or D$_N$(4000). If the D16 profiles accurately represent the metallicity of the gas, then it is not clear from our analysis which physical property may act to cause the dips. Therefore, caution should be taken when interpreting features in metallicity gradients, as we cannot rule out that the dips are caused by diagnostics breaking down under certain conditions.

The variation in the absolute and relative metallicities given by different metallicity diagnostics is widely recognised and there has been a large amount of work already undertaken within this area, using both stacked spectra and spectra from individual H\textsc{II} regions \citep[e.g.][]{Kewley2008, Stasinska2010, Curti2017, Zhang2017, Maiolino2019, Asari2020, Teimoorinia2021}. The differences observed likely reflect the varying biases, or dependencies on secondary parameters, that strong line metallicity diagnostics have; one of the challenges is to find representative samples that can be used to further investigate the impact of changing environmental conditions on the strong line diagnostics. Technological advancements such as the development of more sensitive, higher resolution data such as that returned by MUSE \citep{Bacon2010}, and by JWST in the future, create new opportunities for investigating the origin of the discrepancies observed between the various diagnostics. Ultimately, in order to conclusively determine how various galaxy properties and evolutionary stages impact the observed distribution of metals, we will need sufficiently sensitive observations to measure radial profiles using T$\sub{e}$-based diagnostics.

\subbib

\section{Conclusions}
\label{sect:Conclusions}

Central dips in the metallicity profiles within galaxies have been observed using spatially-resolved IFU data, but the cause of these dips has not yet been conclusively determined.

In this work we first investigated whether there was any evidence of the dips being explained as an artefact introduced by the strong line diagnostics used to determine the metallicity. To do this, we used a sample of galaxies from the MaNGA survey, using four different strong line diagnostics with different known biases to produce radial metallicity profiles. We grouped the galaxies into six different categories according to the shape of the returned profiles, finding large deviations in the shapes of the profiles produced when different diagnostics are used. Stacking spectra from across the sample of galaxies, we compared the results of the strong line diagnostics to those when using T$\sub{e}$-based methods, finding that the D16 diagnostic appeared to underestimate metallicity at 12+log(O/H)$\lesssim 8.4$ across a range of ionisation parameters.  However, we could find no clear evidence of these dips being caused by a changing relationship between metallicity and ionisation parameter in the central regions of galaxies, from that assumed for the corresponding diagnostic. 

We therefore explored whether galaxies exhibiting a dip showed any separation from the rest of the sample in terms of other physical galaxy properties, where we considered both global and spatially-resolved properties. When the O3N2 diagnostic was used to categorise the galaxy metallicity profiles, we found that galaxies with a dip had higher stellar masses and sSFRs compared to the rest of the sample, and showed both lower H$\alpha$EW, and higher central D$_N$(4000) values in average radial profiles. This may therefore indicate that central quenching is related to the presence of dips. However, we found no compelling evidence in the form of higher \textit{B/T} ratios, higher prevalence of outflows, or from the \cite{Wang2020} SFR change parameter, to support this link between observed metallicity dips in the central regions, and quenching occurring within the galaxy. Furthermore, these differences in H$\alpha$EW and D$_N$(4000) profiles observed in galaxies with a central metallicity dip when using the O3N2 diagnostic were not observed when using the D16 metallicity diagnostic to categorise the shape of metallicity profiles. Producing more accurate radial metallicity profiles by using the T$\sub{e}$-based diagnostics would provide greater insight into the differences arising from using the various strong line diagnostics. With the next generation of extremely large telescopes, sufficiently sensitive observations will be available, making these measurements possible.

\section*{Acknowledgements}


We greatly appreciate the MPA-JHU group for making their catalogue public, and C. Avery for sharing their data on the prevalence of outflows within the galaxies in our sample. This research made use of Astropy\footnote{http://www.astropy.org}, a community-developed core Python package for Astronomy \citep{astropy:2013, astropy:2018}. 

This work makes use of data from SDSS-IV. Funding for the Sloan Digital Sky Survey IV has been provided by the Alfred P. Sloan Foundation, the U.S. Department of Energy Office of Science, and the Participating Institutions. SDSS acknowledges support and resources from the Center for High-Performance Computing at the University of Utah. The SDSS web site is www.sdss.org.

SDSS is managed by the Astrophysical Research Consortium for the Participating Institutions of the SDSS Collaboration. 

\section*{Data Availability}


The MaNGA data used in this work was released as part of SDSS DR15 \citep{Aguado2019}, and is publicly available at \url{https://www.sdss.org/dr15/manga/manga-data/}. The MPA-JHU catalogue is available at \url{wwwmpa.mpa-garching.mpg.de/SDSS/DR7/}.


\bibliography{PhD,Additional_bibliography} 



\appendix

\section{Comparing Strong Line to $T\sub{e}$-based Metallicity Diagnostics}
\label{sect:Appendix_D16}

\begin{figure}
    \centering
    \includegraphics[width=1\columnwidth]{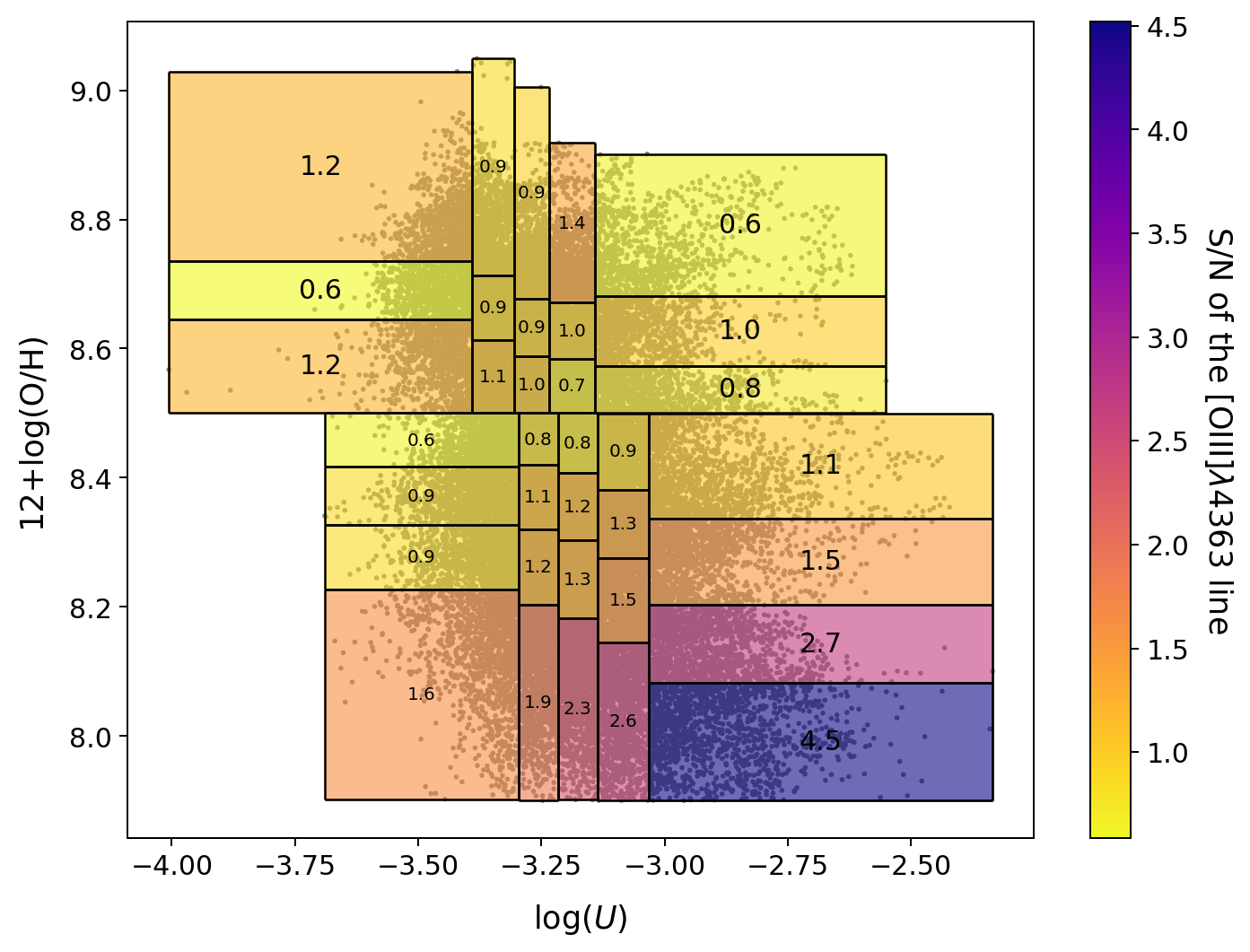}
    \caption{As for Fig.~\ref{fig:OIIISN}, the distribution of 12+log(O/H), as measured by the D16 diagnostic, and $\log(U)$, as measured using the S-based diagnostic presented in \citet{Mingozzi2020} is shown as the grey points. The bin limits are overlaid as the black lines, and the S/N of the \oiiiauroral line as measured using the stacked spectrum from each bin is shown as the colour of the bins, and also as overlaid text.}
    \label{fig:Appendix_D16OIIISN}
\end{figure}

As the \citeauthor{Dopita2016} (2016, D16) metallicity diagnostic is expected to be relatively independent of ionisation parameter, and because of the large $\log(U)$-dependent discrepancies observed in Fig.~\ref{fig:BinvMeasured}, we tested whether using the D16 diagnostic when determining the binning scheme for the stacked spectra would return different results to those presented in Section~\ref{sect:StackedSpectra}. Fig.~\ref{fig:Appendix_D16OIIISN} shows the binning scheme when using the D16 diagnostic to determine metallicity, following the same method as described in Section~\ref{sect:logU}. The points above and below a metallicity of 8.5 were again binned separately, with $\sim$1070 spaxels in each bin above 8.5, and 1488 spaxels in each lower bin. 

\begin{figure*}
    \centering
    \includegraphics[width=0.8\textwidth]{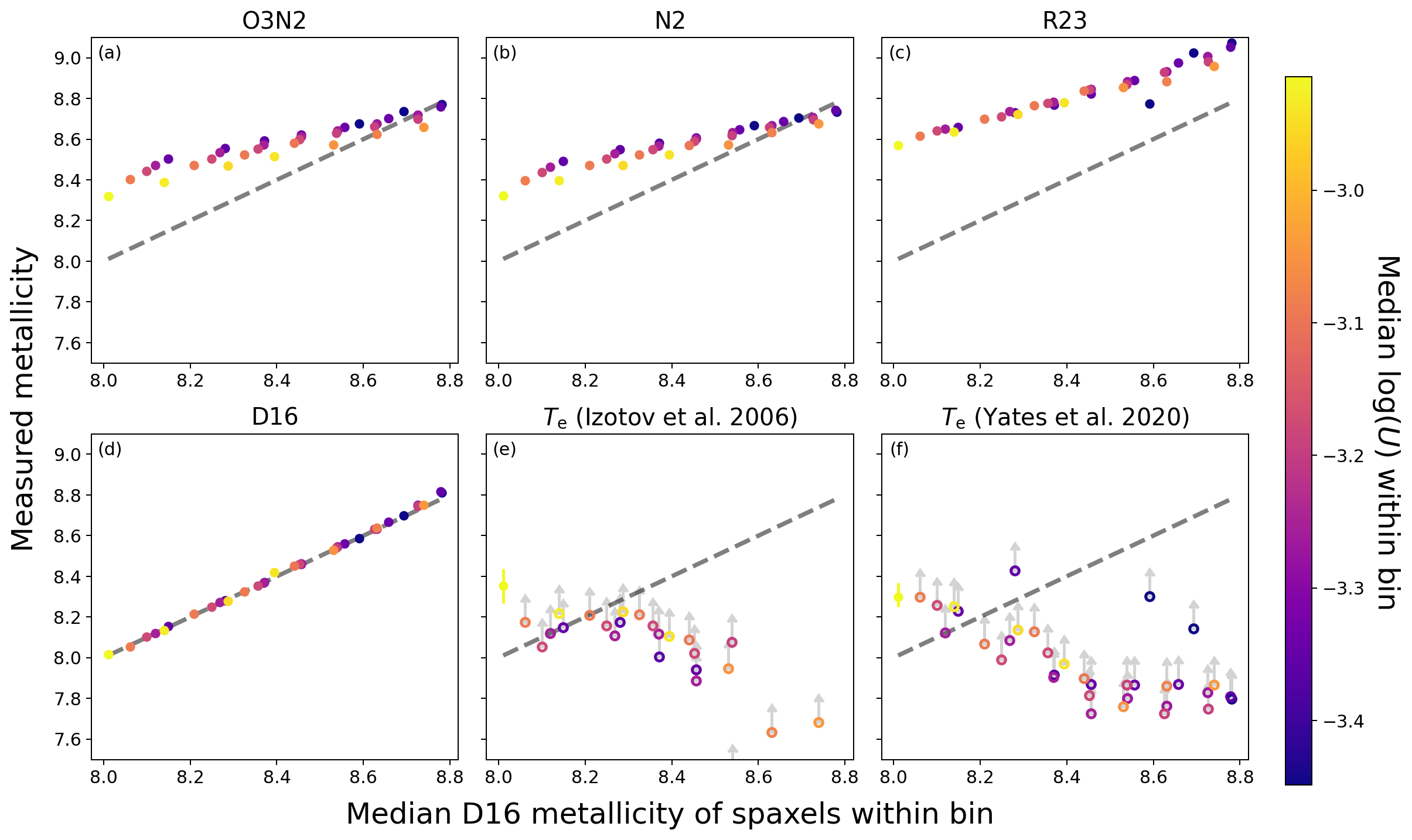}
    \caption{As for Fig.~\ref{fig:BinvMeasured}, the metallicity value measured from each stacked spectrum using different diagnostics is compared to the median value of D16 metallicity for the spaxels within each bin.}
    \label{fig:Appendix_D16BinvMeasured}
\end{figure*}

The spaxels were stacked following the same method as in Section~\ref{sect:logU}, and the S/N of the \oiiiauroral line measured within each stack can be seen in Fig.~\ref{fig:Appendix_D16OIIISN}. Comparing the median metallicity within each bin to the metallicity inferred from the resulting stacked spectrum in Fig.~\ref{fig:Appendix_D16BinvMeasured}, it does appear to remove the $\log(U)$-dependence of the discrepancies for the D16 and R23 diagnostics (panels c and d), when using the D16 diagnostic to determine the bins. However, there still appears to be discrepancies between the D16 and $T\sub{e}$-based methods, for the single bin where the \oiiiauroral could be measured, with D16 under-predicting the metallicity compared to the $T\sub{e}$-based methods (panels e and f).

\section{Global and Spatially-Resolved Properties Using D16 Categorisation}
\label{Appendix:D16}

\begin{figure*}
\centering
\begin{subfigure}[t]{.5\textwidth}
  \centering
  \includegraphics[width=.95\linewidth]{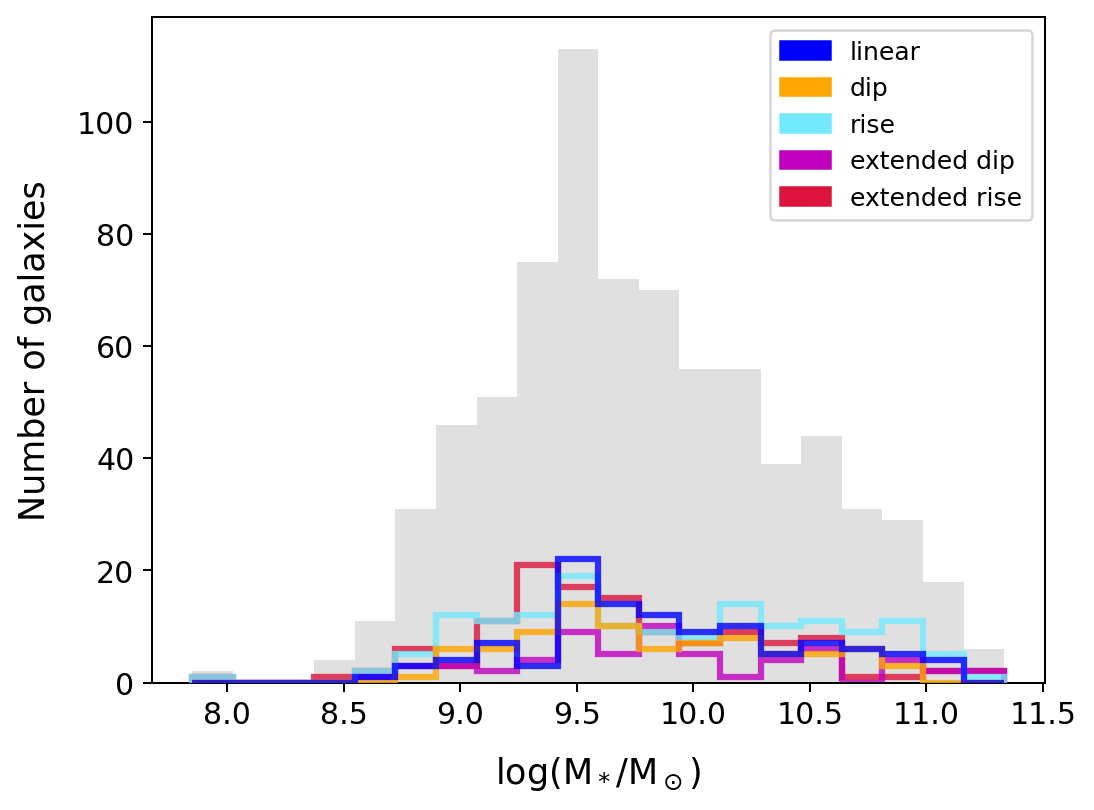}
\end{subfigure}%
\begin{subfigure}[t]{.5\textwidth}
  \centering
  \includegraphics[width=.95\linewidth]{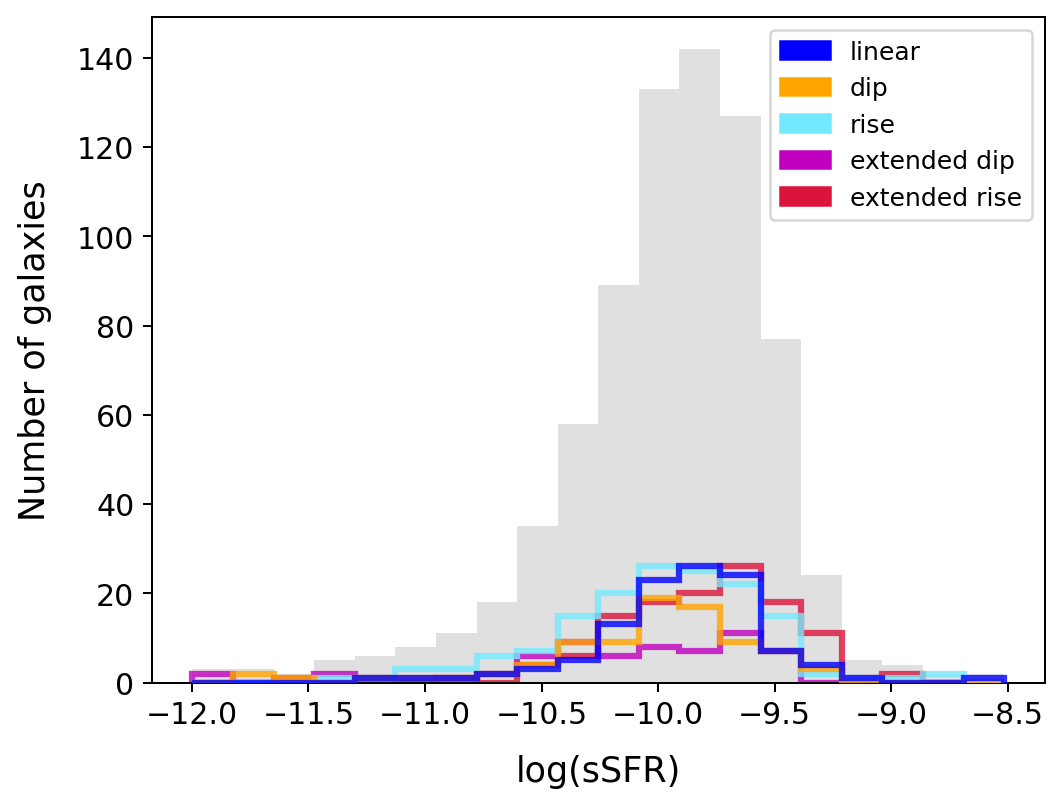}
\end{subfigure}
\caption{As for Fig.~\ref{fig:global_logMstar_sSFR}, the global stellar mass and sSFR are compared for the different D16 metallicity profile categories. There is no longer a clear shift of galaxies with a central dip having higher masses and lower sSFR.}
\label{fig:Appendix_D16_logMlogSFR}
\end{figure*}

\begin{figure*}
    \centering
    \includegraphics[width=0.8\textwidth]{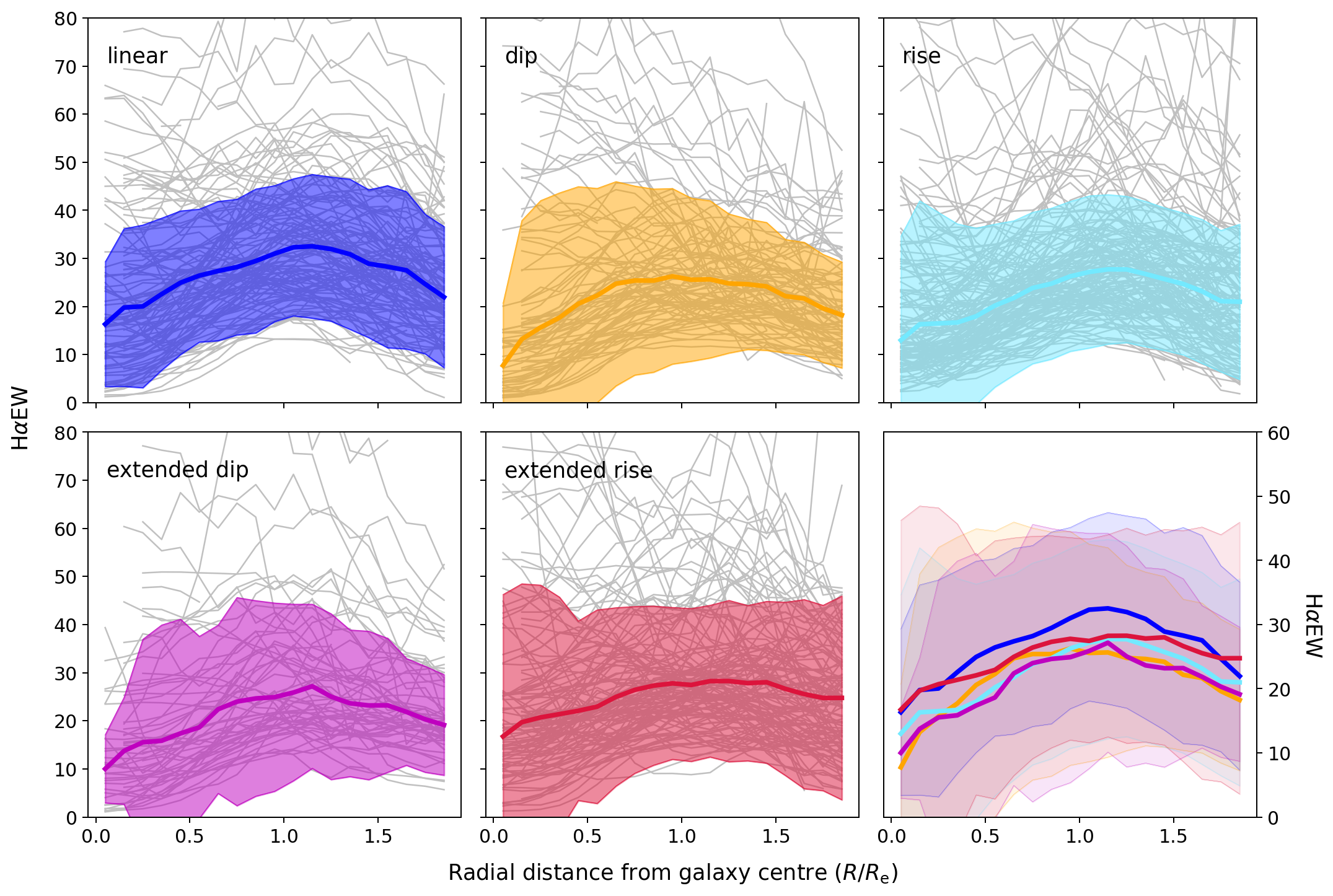}
    \caption{As for Fig.~\ref{fig:avgHaEW}, the average radial H$\alpha$EW are plotted, using D16 to categorise the radial metallicity profiles. There is no longer a clear separation of the galaxies with a central or extended dip having lower average H$\alpha$EW values.}
    \label{fig:Appendix_D16HaEW}
\end{figure*}

\begin{figure*}
    \centering
    \includegraphics[width=0.8\textwidth]{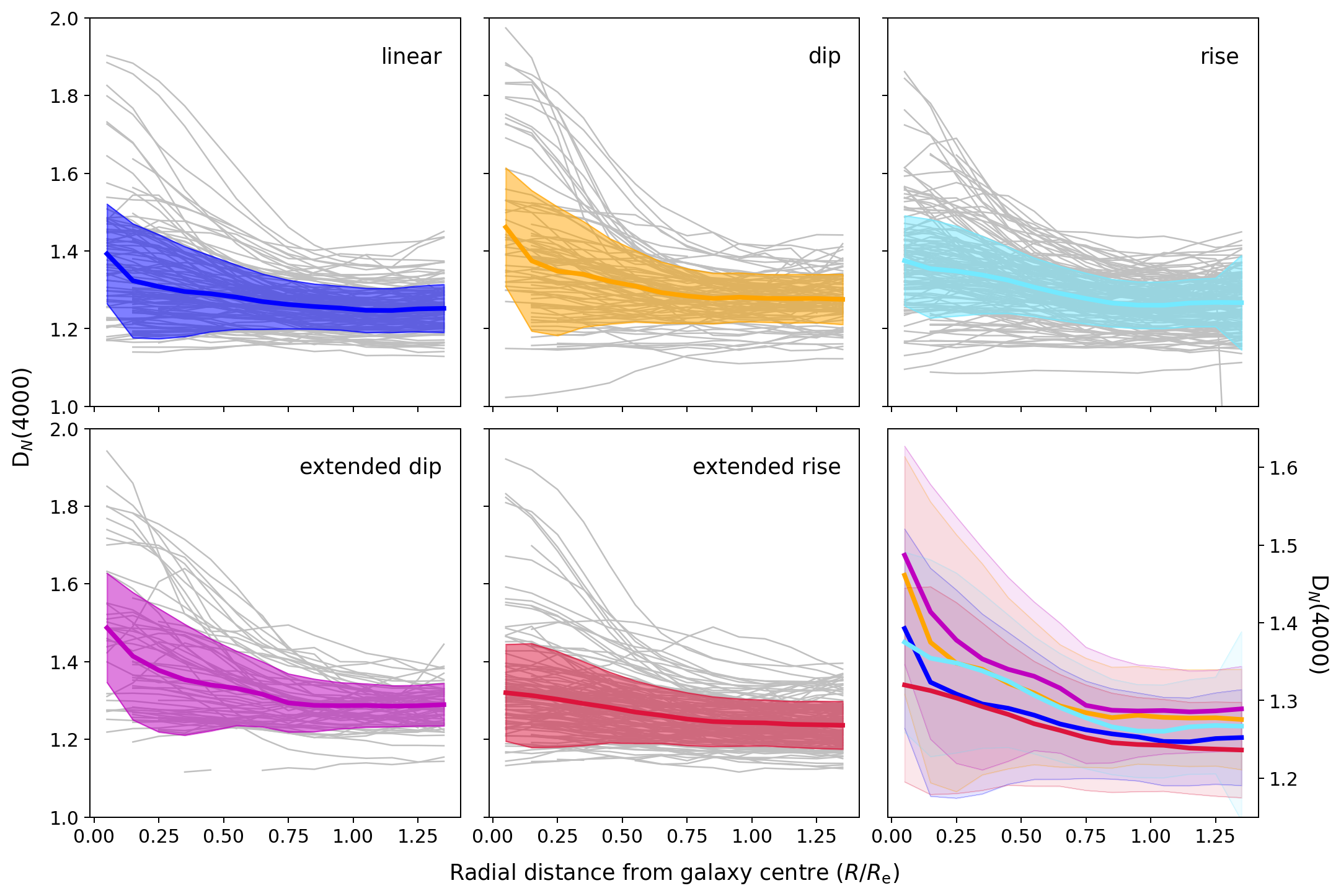}
    \caption{As for Fig.~\ref{fig:avgDN4000}, showing the average radial D$_N$(4000) profiles, using D16 for categorisation of the galaxies. The galaxies with high central D$_N$(4000) values are no longer almost exclusively falling within the dip or extended dip categories, and therefore there is again less of a clear separation between the average profile for these categories, compared to the rest of the sample.}
    \label{fig:Appendix_D16Dn4000}
\end{figure*}

The results presented in Section~\ref{sect:GalaxyProperties} are based on using the O3N2 diagnostic to determine the categorisation of the metallicity profiles. As large differences in the categorisation are observed when different metallicity diagnostics are used (Figs.~\ref{fig:ComparingGradsouter} \& \ref{fig:ComparingGradsinner}), we investigated the impact of using the categorisations of the D16 metallicity profiles instead. D16 was chosen as it has the largest differences with O3N2 in the results presented in Section \ref{sect:Results}. The distribution of global stellar mass and star formation for the galaxies, separated out by the categorisation of the D16 metallicity profiles is shown in Fig. \ref{fig:Appendix_D16_logMlogSFR}. Here it is clear that, when the D16 diagnostic is used, there is no longer a shift towards higher mass and lower star formation rate, as had been observed in Fig.~\ref{fig:global_logMstar_sSFR}. Similarly, in Figs. \ref{fig:Appendix_D16HaEW} and \ref{fig:Appendix_D16Dn4000}, there is much less of a separation of the galaxies categorised as dip and extended dip towards lower H$\alpha$EW values, and the galaxies with high values of D$_N$(4000) in the centres are no longer almost exclusively contained between the dip and extended dip categories.

\section{$\Delta\alpha$ vs global and spatially resolved properties}
\label{Appendix:DeltaSlope}

\begin{figure*}
\centering
\begin{subfigure}[t]{.5\textwidth}
  \centering
  \includegraphics[width=.95\linewidth]{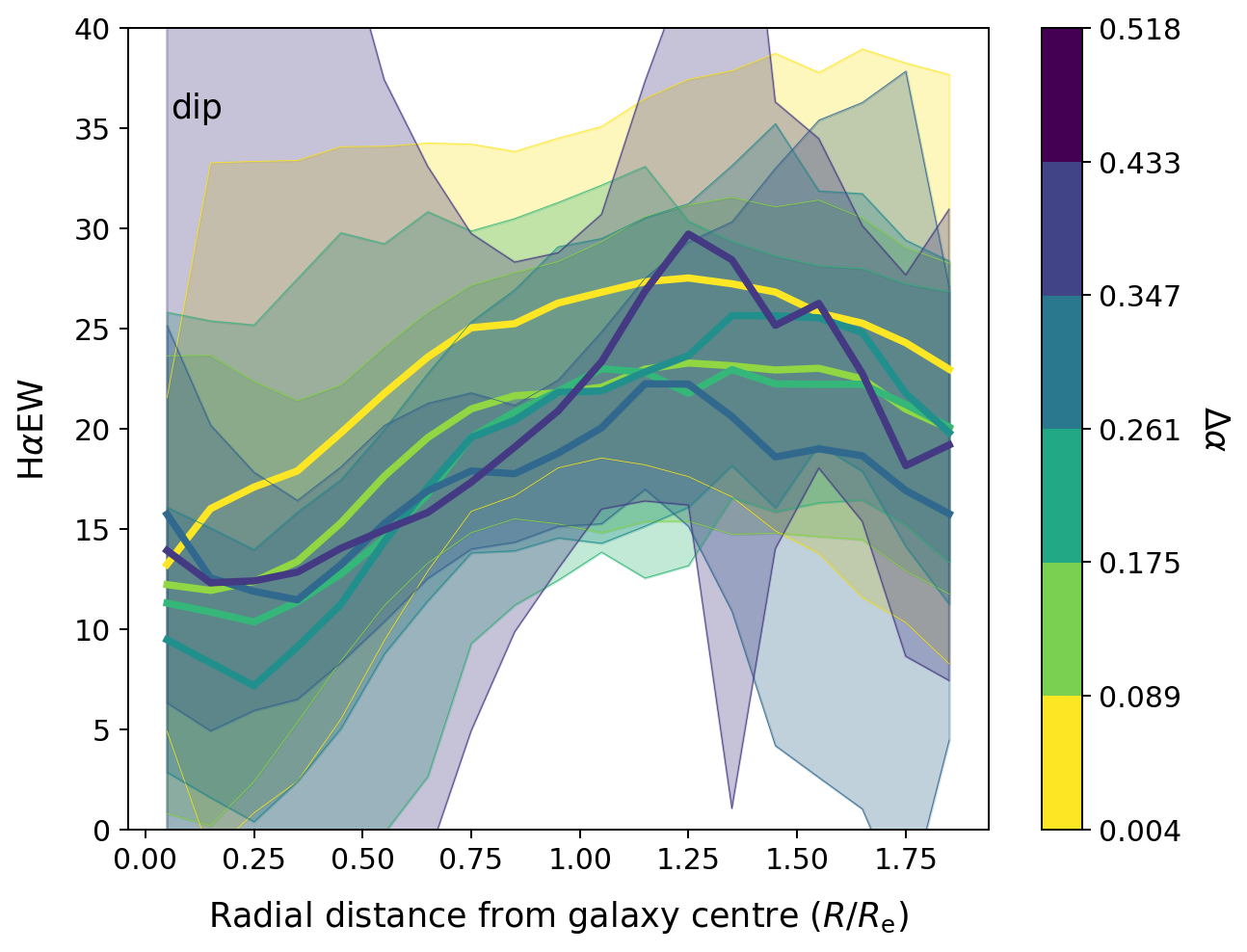}
\end{subfigure}%
\begin{subfigure}[t]{.5\textwidth}
  \centering
  \includegraphics[width=.95\linewidth]{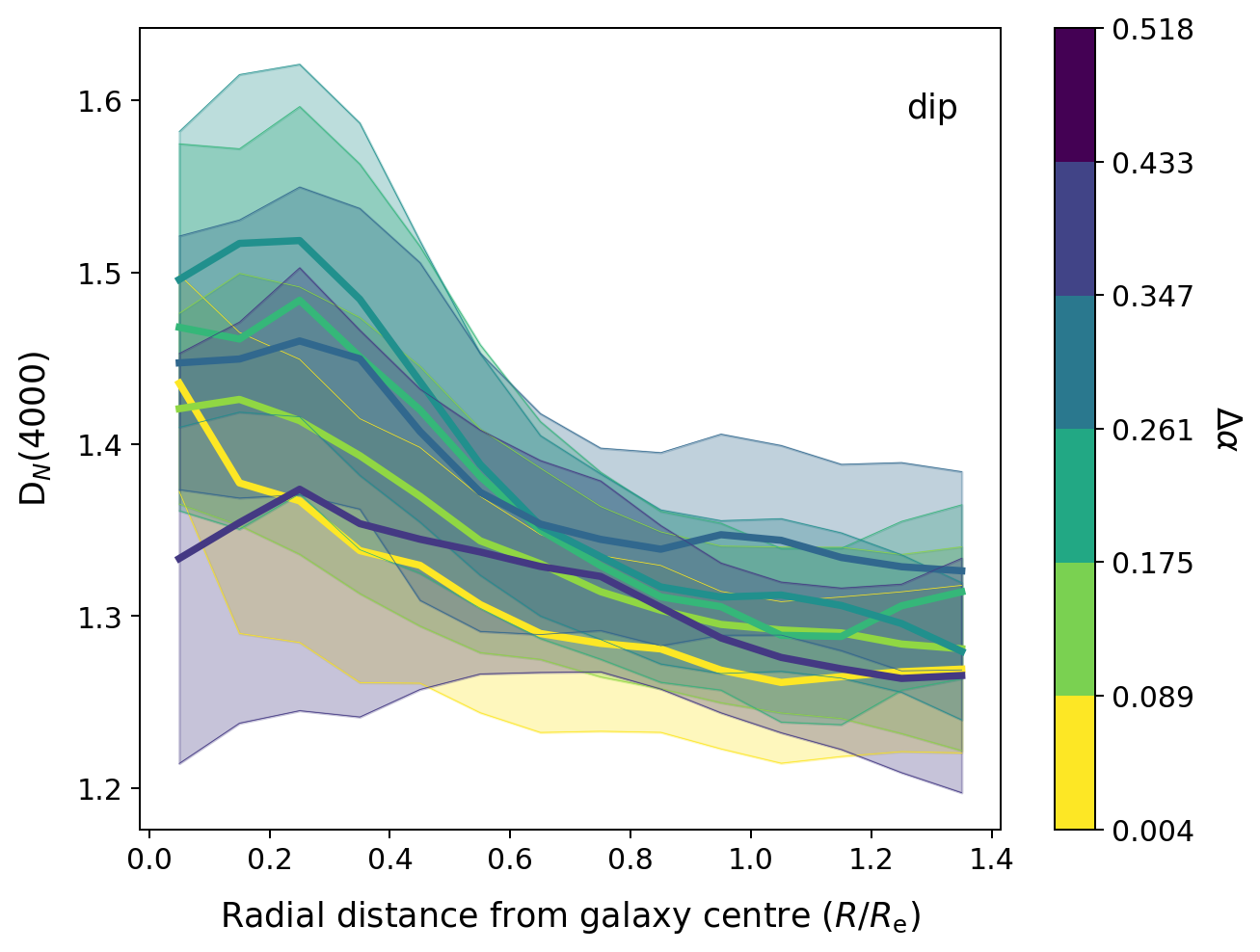}
\end{subfigure}
\caption{Average radial profiles, taken as the median of the weighted mean profiles from individual galaxies, binned by $\Delta\alpha$. \textit{Left:} Average H$\alpha$EW profiles, \textit{Right:} Average D$_N$(4000) profiles. }
\label{fig:Appendix_HaEWDN4000avg}
\end{figure*}

Following a similar method to that used for producing the average radial profiles presented in Section~\ref{subsect:AvgFittedProfiles}, radial profiles of H$\alpha$EW and D$_N$(4000) were created for each galaxy, using a weighted mean. The galaxies categorised as having a dip in their O3N2 profile were then binned by the strength of the dip, $\Delta\alpha$, defined as $\alpha_{in}$ - $\alpha_{out}$. Median values of these individual profiles were then taken for galaxies within each bin, and these averaged profiles can be seen in Fig.~\ref{fig:Appendix_HaEWDN4000avg}. There does seem to be a slight correlation between $\Delta\alpha$ and the normalisation of the H$\alpha$EW, with galaxies showing stronger dips in metallicity having slightly lower normalisations, with the exception of the highest $\Delta\alpha$ bins. For the D$_N$(4000) profiles, there is a slight trend towards higher normalisations for galaxies with the largest changes in slope, although again this trend is not followed for the highest $\Delta\alpha$ bins. The implications of this are discussed further in Section~\ref{sect:Discussion}.

\subbib


\bsp	
\label{lastpage}
\end{document}